\documentclass[     twocolumn,    %
showpacs,eqsecnum,prb,amsmath,amssymb]{revtex4}
\usepackage{graphicx}
\usepackage{dcolumn}
\usepackage{bm}
\begin{document}
\title{%
Fermi liquid in the Hubbard Model with an electron reservoir: \\
Normal state of cuprate superconductors
}
\author{Fusayoshi J. Ohkawa}
\affiliation{Department of Physics, Faculty of Science, 
Hokkaido University, Sapporo 060-0810, Japan}
\email{fohkawa@phys.sci.hokudai.ac.jp}
\received{July 1, 2008}
\begin{abstract}  
It is proved that the ground state under the supreme single-site approximation (S$^3$A), the dynamical mean-field theory (DMFT), or the dynamical coherent potential approximation (DCPA) is the normal Fermi liquid in the presence of an infinitesimally weak hybridization with an electron reservoir, except for the just half filling of electrons and the infinite on-site repulsion.
In the strong-coupling regime, in particular, the Fermi liquid is stabilized under S$^3$A, DMFT, or DCPA by the Kondo effect, which stabilizes a local singlet on each unit cell, and is further stabilized beyond it by the Fock-type term of the superexchange interaction or a resonating valence bond (RVB) mechanism, which stabilizes a local singlet on each pair of nearest neighbors.
The Fermi liquid is a relevant {\it normal} state to study possible lower-temperature phases or the true ground state. 
It is proposed that the Fermi liquid stabilized by the Kondo effect and the RVB mechanism is the normal state of cuprate high-temperature superconductors. 
\end{abstract}
%
%
\pacs{71.10.-w, 71.10.Ay, 71.27.+a, 74.20.-z}
 %
%
\maketitle
\section{Introduction}
\label{SecIntroduction}
High temperature (high-$T_{\rm c}$) superconductivity in cuprates 
is an interesting and important issue in solid-state physics. 
\cite{bednortz,RevD-wave,RevScience,RevStripe,lee,RevScanning}
Parent cuprates are insulators, which show antiferromagnetism at low temperatures. When {\it holes} or electrons are doped, they become metals, which show exotic properties. \cite{RevD-wave,RevScience,RevStripe,lee,RevScanning} High-$T_{\rm c}$ superconductivity occurs in exotic metals in the vicinity of the Mott metal-insulator \mbox{(M-I)} transition, which is also an interesting and important issue. \cite{mott,tokura,PhyToday,lee} 
In order to resolve the issue on the mechanism of high-$T_{\rm c}$ superconductivity, 
it should be clarified whether the {\it normal} state above $T_{\rm c}$ is an exotic Fermi liquid (FL), such as the resonating valence bond (RVB) state, \cite{RVB} or the conventional or normal FL. First of all, the nature of electron correlations in the vicinity of the Mott transition should be clarified.

The Hubbard model is an effective Hamiltonian for the Mott  transition.
According to Hubbard's theory,\cite{Hubbard1,Hubbard2} 
when $U\agt W$, with $U$ the on-site repulsion and $W$ the band-width, the band splits into two subbands, i.e.,
the Hubbard gap opens between the upper Hubbard band (UHB) and the lower Hubbard band (LHB). 
When $n=1$, with $n$ the electron density per unit cell, the ground state is a prototype of the Mott insulator, which
seems to be an abnormal insulator characterized by the ground-state entropy diverging in the thermodynamic limit; \cite{Hubbard3} the insulator for $n=1$ and $U/W=+\infty$ is the typical Mott insulator, whose entropy is $k_{\rm B}\ln2$ per unit cell. 
When $U\alt W$ or $n\ne 1$, the ground state is expected to be a metal since the density of states (DOS) at the chemical potential is nonzero; the Fermi surface (FS) cannot be defined within Hubbard's theory.

According to Gutzwiller's theory,\cite{Gutzwiller1,Gutzwiller2,Gutzwiller3} together with the FL theory, \cite{Luttinger1,Luttinger2} the FS is defined in the quasi-particle band, which is called the Gutzwiller band in this paper. 
According to Brinkman and Rice's theory, \cite{brinkman} however, when $n=1$ the Gutzwiller band vanishes and an \mbox{M-I} transition occurs at $U= U_{\rm BR}$, with $U_{\rm BR} \simeq W$. 
The specific heat coefficient is diverging as $U\rightarrow U_{\rm BR}-0$, which implies that the ground state is the Mott insulator or the abnormal insulator when $n=1$ and $U\ge U_{\rm BR}$. 
The ground state is the FL when $n\ne 1$ or $U< U_{\rm BR}$.

When $U$ is large but is still finite, the perturbative process that gives the superexchange interaction never vanishes,\cite{andersonJ} so that matrix elements must be nonzero among the degenerate ground states if their configurations are close to each other.
It should be critically examined whether the abnormal insulator is really stable for finite $U$ or the third law of thermodynamic is really broken for finite $U$,
but within the restricted Hilbert subspace where no order parameter exists; it is obvious that the abnormal insulator for finite $U$ is unstable against an antiferromagnetic state in the whole Hilbert space.

One may speculate that DOS has a three-peak structure, with the Gutzwiller band between UHB and LHB, in a metallic phase of $n\simeq 1$ and $U\agt W$.
Hubbard's and Gutzwiller's theories are under the single-site approximation (SSA). According to another SSA theory, \cite{OhkawaSlave} the Gutzwiller band appears at the top of LHB for $n<1$, which implies that it appears at the bottom of UHB for $n>1$. 
The SSA that considers all the single-site terms is rigorous for $d\rightarrow +\infty$ within the restricted Hilbert subspace, \cite{Metzner,Muller-H1,Muller-H2,Janis,comAttractive} 
with $d$ being the spatial dimensionality.
The SSA is called the supreme single-site approximation (S$^3$A) in this paper.
The S$^3$A is reduced to solving the Anderson model, \cite{Mapping-1,Mapping-2,Mapping-3,georges} which is an effective Hamiltonian for the Kondo effect. 
The three-peak structure corresponds to that in the Anderson model, with the Kondo peak between two subpeaks. 
The Kondo effect has relevance to electron correlations in the vicinity of the Mott transition.
The S$^3$A is also formulated as the dynamical mean-field theory\cite{georges,RevMod,kotliar,PhyToday} (DMFT) and the dynamical coherent potential approximation \cite{dpca} (DCPA).

A Kondo-lattice theory \cite{Mapping-1,Mapping-2,Mapping-3} 
and a cluster DMFT (CDMFT) \cite{cellDMFT1,cellDMFT2,cellDMFT3,cellDMFT4} 
are proposed beyond S$^3$A or DMFT. 
In the Kondo-lattice theory, it is assumed that the ground state under S$^3$A or DMFT is the FL, i.e., the Gutzwiller band and the FS exist or they survive even if the Hubbard gap opens. 
\cite{Mapping-1,Mapping-2,Mapping-3} 
Numerical results of S$^3$A or DMFT and those of CDMFT give an indication that, when $n\simeq 1$ and $U/W\agt 1$, the Gutzwiller band vanishes and the ground state is an insulator,  \cite{RevMod,kotliar,PhyToday,cellDMFT1,cellDMFT2,cellDMFT3,cellDMFT4} i.e., the Mott insulator or a spin liquid.
It is a crucial issue which is the ground state under S$^3$A, DMFT, or DCPA, or within the restricted Hilbert subspace, the normal FL, an exotic FL, the Mott insulator, or a spin liquid. This issue is related to that on the normal state of cuprate superconductors.

The $t$-$J$ model is an effective Hamiltonians for cuprate superconductors, which are anisotropic quasi-two dimensional oxides composed of CuO$_2$ planes. It is derived from the Hubbard model\cite{hirsch} and the $d$-$p$ model, \cite{rice} which considers $d$ orbits on Cu ions and $p$ orbits on O ions on CuO$_2$ planes. When $n=1$, the $t$-$J$ model is reduced to the Heisenberg model. 
Anderson proposes the RVB theory of high-$T_{\rm c}$ superconductivity in the $t$-$J$ model. \cite{RVB} A parent state is a spin liquid or the RVB state \cite{fazekas} in the Heisenberg model, rather than the Mott insulator. When {\it holes} or electrons are doped, it becomes a metallic RVB state, which is the normal state in the RVB theory. \cite{RVB} The RVB state, which is an insulator or a metal, is stabilized by the formation of a resonating valence bond due to the superexchange interaction.
In the mean-field RVB theory,\cite{Plain-vanilla} the metallic RVB state is stabilized by the Fock-type exchange interaction. 
According to the Kondo-lattice theory for the $t$-$J$ model, \cite{phase-diagram} the FL is stabilized under S$^3$A, DMFT, DCPA by the Kondo effect and is further stabilized beyond it by the Fock-type exchange interaction. The RVB state and the FL are similar to each other. Similarities and differences between the RVB state and the FL should be clarified. 

This paper is organized as follows: 
In Sec.~\ref{SecSSA}, it is proved that 
the ground state of the Hubbard model under S$^3$A, DMFT, or DCPA is the FL when an infinitesimally weak hybridization with an electron reservoir exists, except for $n=1$ and $U/W=+\infty$.
In Sec.~\ref{SecBeyondSSA}, it is shown that the FL, which is further stabilized by the Fock-type exchange interaction beyond S$^3$A, DMFT, or DCPA, is eventually unstable against at least a magnetic or superconducting state in two dimensions and higher.
In Sec.~\ref{SecDiscussion}, several issues are discussed:
impossibility of the Mott insulator for finite $U$, the normal state of cuprate superconductors, and so on.
Conclusion is given in Sec.~\ref{SecConclusion}. 
An inequality, which is used in the proof in Sec.~\ref{SecSSA}, is proved in Appendix~\ref{SecProof}.
In Appendix~\ref{SecSpin}, the FL theory for a spin liquid in the Heisenberg model is developed and it is proposed that the insulating and metallic RVB states are the spin liquid and the FL, respectively.
In Appendix~\ref{SecAppendixGutzwiller}, on the basis of an analysis that single-particle excitations are different between the presence and absence of an electron reservoir,
it is shown that the well-known physical picture for the Mott transition, which is one in the absence of an electron reservoir, 
is never relevant to explain the Mott transition.

\section{Fermi liquid under S$^3$A or DMFT}\label{SecSSA}
\subsection{Perturbation from an electron reservoir}
In this paper, a reservoir is explicitly considered:
\begin{equation}\label{EqGrandH}
\bar{\cal H} = {\cal H} + {\cal H}_{\rm res} + \lambda{\cal V}-
\mu \bar{\cal N}.
\end{equation}
The first term is the Hubbard model defined by 
\begin{equation}\label{EqHubbard}
{\cal H} = \epsilon_a
\sum_{i\sigma}n_{i\sigma} + \sum_{i\ne j,\sigma} t_{ij}
a_{i\sigma}^\dag a_{j\sigma}
+U \sum_{i} n_{i\uparrow} n_{i\downarrow},
\end{equation}
with $n_{i\sigma}= a_{i\sigma}^\dag a_{i\sigma}$, $\epsilon_a$ the band center, $t_{ij}$ transfer integrals, and $U$ the on-site repulsion. 
The dispersion relation of electrons is defined by 
\begin{equation}\label{EqBareDispersion}
E({\bf k}) = \epsilon_a +
\frac1{N_{\rm c}} \sum_{i\ne j} t_{ij} \exp\left[i{\bf k}\cdot
\left({\bf R}_i-{\bf R}_j\right)\right] ,
\end{equation}
with $N_{\rm c}$ the number of unit cells and ${\bf R}_i$ the position of the
$i\hskip1pt$th unit cell. DOS is defined by
\begin{equation}\label{EqRho0}
D(\varepsilon) = \frac1{N_{\rm c}} \sum_{\bf k}
\delta[\varepsilon - E({\bf k})] .
\end{equation}
The band-width of $E({\bf k})$ or $D(\varepsilon)$ is denoted by $W$. The electron density $n$ per unit cell in the Hubbard model is defined by $n=\left<{\cal N}\right>/N_{\rm c}$, with
\begin{equation}\label{EqNumber}
{\cal N} = \sum_{i\sigma} n_{i\sigma}.
\end{equation}
The chemical potential for $U=0$ and $T=0$~K, which is denoted by $\mu_0(n)$, is defined by
\begin{equation}\label{EqN0}
n = 
%
\frac{2}{N_{\rm c}}\sum_{{\bf k}} 
\theta\bigl( [ \mu_0(n) - E({\bf k}) ]/W \bigr) , 
\end{equation}
with $\theta(x)$ the Heaviside function, i.e., $\theta(x)=0$ for $x<0$ and $\theta(x)=1$ for $x\ge0$.
It is assumed that $D(\varepsilon)$ is nonzero, continuous, and finite at least at $\varepsilon=\mu_0(n)$.
The second term stands for the reservoir: 
\begin{equation}
{\cal H}_{\rm res} = \sum_{{\bf k}\sigma}
E_b({\bf k}) b_{{\bf k}\sigma}^\dag b_{{\bf k}\sigma}.
\end{equation}
The third term is a {\it hybridization} term between the Hubbard model and the reservoir:
\begin{equation}
\lambda{\cal V} = \lambda \sum_{i \in {\cal R}}\left[
v_i a_{i\sigma}^\dag b_{i \sigma}
+v_i^* b_{i\sigma}^\dag a_{i \sigma} \right] ,
\end{equation}
with the summation over $i$ being over ${\cal R}$ of randomly distributed hybridization sites, and
\begin{equation}\label{EqBOp}
b_{i\sigma} = \frac1{\sqrt{N_{\rm c}}}\sum_{\bf k} b_{{\bf k}\sigma}
e^{i{\bf k}\cdot{\bf R}_i}.
\end{equation} 
It is assumed that $\lambda$ is a nonzero but infinitesimally small numerical constant, which is denoted by $\lambda=\pm 0$.
Unless $\lambda= 0$, the electron number ${\cal N}$ in the Hubbard model is a non-conserved quantity.
It is assumed that
$\left<\hskip-2pt\left<\hskip1pt v_i
\hskip1pt\right>\hskip-2pt\right> =
\left<\hskip-2pt\left<\hskip1pt  v_i^*
\hskip1pt\right>\hskip-2pt\right> =0$ 
and 
$\left<\hskip-3pt\left<\hskip1pt v_i v_j^*
\hskip1pt\right>\hskip-3pt\right> =
\delta_{ij} n_{\rm h}|v|^2$, 
%
with $\left<\hskip-2pt\left<\hskip1pt\cdots\hskip1pt\right>\hskip-2pt\right>$ standing for the ensemble average for ${\cal R}$ and $n_{\rm h}$ the density of hybridization sites per unit cell.
In the last term, 
$\mu$ is the chemical potential, and
\begin{equation}
\bar{\cal N} = \sum_{i\sigma}\left(a_{i\sigma}^\dag a_{i\sigma}
+b_{i\sigma}^\dag b_{i\sigma} \right).
\end{equation}

The Green function for electrons in the Hubbard model with $U=0$ averaged over the ensemble is given by
\begin{equation}\label{EqG0}
G_{\sigma}^{(0)}(i\varepsilon_n,{\bf k})=
\frac1{i\varepsilon_n +\mu - E({\bf k}) 
+ i \lambda^2\Gamma (i\varepsilon_n) }.
\end{equation}
%
When $\lambda=\pm 0$, the second-order perturbation is accurate enough to treat scatterings from the random hybridization, so that 
\begin{equation}
\Gamma (i\varepsilon_n) = 
i n_{\rm h}|v|^2\frac1{N_{\rm c}}\sum_{\bf k}
\frac1{i\varepsilon_n +\mu -E_b({\bf k})}.
\end{equation}
It is assumed that no gap opens in the reservoir or that $\Gamma (\varepsilon+i0)$ is continuous at $\varepsilon=0$ and
\begin{equation}\label{EqFS-P1}
\mbox{Re} \hskip2pt \Gamma(+i0)> 0. 
\end{equation}

For the sake of simplicity,
it is assumed throughout of this paper that the model (\ref{EqGrandH}) is on a lattice in two dimensions and higher and $t_{ij}$ is only nonzero between nearest neighbors $\left<ij\right>$, if nothing is mentioned about dimensionality $d$ or $t_{ij}$; $t_{\left<ij\right>}$ is denoted by $t$, if necessary.

\subsection{Proof of the FL ground state}
\label{SecProofGround}
\subsubsection{Fermi-surface (FS) condition in the Anderson model}
\label{SecFS-condition}
The \mbox{$s$-$d$} model is another effective Hamiltonian for the Kondo effect.
According to Yosida's perturbation theory \cite{yosida}
and Wilson's renormalization-group theory, \cite{wilsonKG} when DOS of the conduction band is nonzero at the chemical potential,
the ground state is a singlet or the normal FL but is exceptionally a doublet for $J_{sd}=0$, with $J_{sd}$ the $s$-$d$ exchange interaction. 

The $s$-$d$ model is derived from the Anderson model: 
\begin{eqnarray}\label{EqAnderson}
{\cal H}_{\rm A} &=&
\sum_{{\bf k}\sigma} E_c({\bf k}) 
c_{{\bf k}\sigma}^\dag c_{{\bf k}\sigma}
+ \epsilon_d \sum_{\sigma}n_{d\sigma}
+ \tilde{U} n_{d\uparrow} n_{d\downarrow}
\nonumber \\ && 
+ \frac1{\sqrt{ N_{\rm A} }} \sum_{{\bf k}\sigma} \left(
V_{\bf k}c_{{\bf k}\sigma}^\dag d_\sigma
+ V_{\bf k}^* d_\sigma^\dag c_{{\bf k}\sigma}\right) ,
\end{eqnarray} 
with $n_{d\sigma}=d_{\sigma}^\dag d_{\sigma}$, $\epsilon_d$ the $d$ electron level, $\tilde{U}$ the on-site repulsion, $V_{\bf k}$ the hybridization matrix, and $N_{\rm A}$ the number of unit cells. 
The hybridization energy is defined by
\begin{equation}
\Delta(\varepsilon) =
\frac{\pi}{N_{\rm A}} \sum_{\bf k} |V_{\bf k}|^2 
\delta\bigl[\varepsilon+\tilde{\mu }- E_c({\bf k})\bigr] ,
\end{equation}
with $\tilde{\mu}$ the chemical potential. The FS of conduction electrons, which is defined by $E_c({\bf k})=\tilde{\mu}$, exists when 
\begin{equation}\label{EqFS}
\Delta(0) > 0 ,
\end{equation}
which is called the FS condition in this paper. The results for the $s$-$d$ model imply that, when the FS condition is satisfied,
the ground state is a singlet or the FL but is exceptionally a doublet for $n_d=1$ and $\tilde{U}/\pi\Delta(0)=+\infty$, with $n_d = \sum_\sigma\left<n_{d\sigma}\right> $ being the density of $d$ electrons. 

According to Bethe-ansatz solutions for the $s$-$d$ model with a constant DOS of the conduction band and the Anderson model with a constant $\Delta(\varepsilon)$, the ground state of either model is the FL except for each exceptional case. \cite{exact1,exact2,exact3,exact4} 
In general, the nature of the ground state depends only on relevant low-energy properties, such as $\Delta(0)$, and high-energy properties renormalize only quantitatively the ground state, as is demonstrated by renormalization-group theories for the $s$-$d$ model.\cite{wilsonKG,poorman} 
The Kondo effect is almost or practically solved.\cite{wilsonKG,exact1,exact2,exact3,exact4} 
The most fundamental assumption of this paper is that when the FS condition is satisfied the ground state of the Anderson model is the FL except for $n_d=1$ and $\tilde{U}/\pi\Delta(0)=+\infty$.

\subsubsection{Mapping to the Anderson model}
\label{SecMapping}
Consider the Hubbard model within the restricted Hilbert subspace where no order parameter exists. The Green function averaged over the ensemble is given by
\begin{equation}\label{EqGreen}
G_{\sigma}(i\varepsilon_n, {\bf k}) =
\frac1{i\varepsilon_n + \mu 
- E({\bf k}) - \Sigma_{\sigma}(i\varepsilon_n, {\bf k}) 
+ i \lambda^2 \Gamma(i\varepsilon_n)}.
\end{equation}
Here, $\Sigma_{\sigma}(i\varepsilon_n, {\bf k})$ is the self-energy, which is divided into single-site $\tilde{\Sigma}_{\sigma}(i\varepsilon_n)$ and multi-site $\Delta \Sigma_{\sigma}(i\varepsilon_n, {\bf k})$:
\begin{equation}
\Sigma_{\sigma}(i\varepsilon_n, {\bf k})
= \tilde{\Sigma}_{\sigma}(i\varepsilon_n)
+ \Delta \Sigma_{\sigma}(i\varepsilon_n, {\bf k}).
\end{equation}
Since any vertex correction due to $\lambda{\cal V}$ and ${\cal H}_{\rm res}$ is $O(\lambda^4)$ or higher, it can be ignored when $\lambda=\pm 0$.
The single-site $\tilde{\Sigma}_{\sigma}(i\varepsilon_n)$ is given by that of the Anderson model provided that the on-site repulsion line and the single-site electron lines in Feynman diagrams are the same as each other between the Hubbard and Anderson models. 
If $\tilde{\Sigma}_{\sigma}(i\varepsilon_n)$ and $\Delta \Sigma_{\sigma}(i\varepsilon_n, {\bf k})$ are obtained, 
the single-site Green function of the Hubbard model is given by
\begin{equation}\label{EqR}
R_{\sigma}(i\varepsilon_n) =
\frac1{N_{\rm c}} \sum_{\bf k} G_{\sigma}(i\varepsilon_n,{\bf k}),
\end{equation}
and that of the Anderson model is given by
\begin{equation}
\tilde{G}_{\sigma}(i\varepsilon_n) =
\frac1{\displaystyle 
i\varepsilon_n + \tilde{\mu}
 - \epsilon_d - \tilde{\Sigma}_{\sigma}(i\varepsilon_n)
- \frac1{\pi} \hskip-2pt \int \hskip-3pt d\varepsilon^\prime 
\frac{\Delta(\varepsilon^\prime)}
{i\varepsilon_n - \varepsilon^\prime} } .
\end{equation}
The condition for the electron lines is given by
$R_{\sigma}(i\varepsilon_n) = \tilde{G}_\sigma (i\varepsilon_n)$
or
\begin{equation}\label{EqMappingCondition1}
R_{\sigma}(\varepsilon+i0) = \tilde{G}_\sigma (\varepsilon+i0),
\end{equation}
which can never be satisfied unless 
\begin{subequations}\label{EqMappingCondition}
\begin{equation}\label{EqMappingConditionL}
\tilde{\mu}-\epsilon_d = \mu - \epsilon_a .
\end{equation}
It follows from (\ref{EqMappingCondition1}) that
\begin{equation}\label{EqMappingCondition2}
\Delta (\varepsilon)= \mbox{Im} \left[
\tilde{\Sigma}_\sigma(\varepsilon+i0) 
+ R_\sigma^{-1}(\varepsilon + i0) \right].
\end{equation}
The condition for the on-site repulsion line is given by
\begin{equation}
\tilde{U} = U .
\end{equation}
\end{subequations}
A problem of calculating the single-site $\tilde{\Sigma}_\sigma(i\varepsilon_n)$ is reduced to that of determining and solving self-consistently the Anderson model to satisfy Eq.~(\ref{EqMappingCondition}), which is called the mapping condition in this paper;
the multi-site $\Delta \Sigma_\sigma(i\varepsilon_n, {\bf k})$ should also be self-consistently calculated with the single-site $\tilde{\Sigma}_\sigma(i\varepsilon_n)$ to satisfy Eq.~(\ref{EqMappingCondition}). 
According to Eq.~(\ref{EqMappingCondition1}),
DOS and the electron density are the same as each other between the two models:
\begin{equation}\label{EqRhoSSA}
\rho(\varepsilon) 
= -\frac1{\pi}\mbox{Im} R_\sigma(\varepsilon+i0)
= -\frac1{\pi}\mbox{Im} \tilde{G}_\sigma(\varepsilon+i0) ,
\end{equation}
and
\begin{equation}
n = n_d = 2\int_{-\infty}^{+\infty} d\varepsilon
f(\varepsilon)\rho(\varepsilon) ,
\end{equation}
with
$f(\varepsilon) = 1/[\exp(\varepsilon/k_{\rm B}T) +1] $
being the Fermi-Dirac function.

In S$^3$A, DMFT, or DCPA, only the single-site $\tilde{\Sigma}_\sigma(i\varepsilon_n)$ is considered, so that 
\begin{equation}\label{EqR-delta1}
R_\sigma(\varepsilon \!+\! i0) =
\hskip-2pt \int \hskip-3pt d \varepsilon^\prime
\frac{D(\varepsilon^\prime) }
{\varepsilon \!+\! \mu \!-\! 
\tilde{\Sigma}_\sigma(\varepsilon \!+\! i0)
\!+\! i \lambda^2 \Gamma (\varepsilon \!+\! i0) \!-\! \varepsilon^\prime}.
\end{equation}
The mapping condition~(\ref{EqMappingCondition2}) is iteratively treated to determine the Anderson model to be solved.
Even if any $\tilde{\Sigma}_{\sigma}(\varepsilon+i0)$ is assumed in Eq.~(\ref{EqMappingCondition2}), \cite{comHubbard} it follows that
%
\begin{equation}\label{EqImp}
\Delta (\varepsilon) \ge\mbox{Re}\hskip2pt
\lambda^2 \Gamma(\varepsilon+i0),
\end{equation}
as is proved in Appendix~\ref{SecProof}.
According to Eqs.~(\ref{EqFS-P1}) and (\ref{EqImp}), the FS condition (\ref{EqFS}) is satisfied in each iterative process. 
Therefore, any self-consistent solution is of the normal FL.
When $\lambda= \pm0$ or ${\cal N}$ is a non-conserved quantity, the ground state under S$^3$A, DMFT, or DCPA, is the normal FL except for $U/W=+\infty$ and $n=1$.

\subsection{Fermi-liquid (FL) relation}
\label{SecFL-Relation}
Consider the mapped Anderson model in the presence of 
an infinitesimally small Zeeman energy $h_Z=g\mu_{\rm B}H$ and an infinitesimally small chemical potential shift $\Delta\mu$. 
The self-energy is expanded at $T=0$~K such that
\begin{eqnarray}\label{EqExpandSelf}
\tilde{\Sigma}_\sigma(\varepsilon + i0) &=&
\tilde{\Sigma}_0(0)
+ \left(1 - 
\tilde{\phi}_\gamma \right)\varepsilon 
+ \left(1 - 
\tilde{\phi}_s \right) \frac1{2}\sigma h_Z
\nonumber \\ &&
+ \left(1 - \tilde{\phi}_c \right) \Delta\mu
+ O\left(\varepsilon^2\right), 
\end{eqnarray}
with $\tilde{\Sigma}_0(0)$, $\tilde{\phi}_\gamma$,
$\tilde{\phi}_s$ and $\tilde{\phi}_c$ all being real and finite.
Since DOS of the Anderson model is the same as that of the Hubbard model, it follows that 
\begin{equation}\label{EqRhoSSA1}
\rho(0) =
\frac{1}{N_{\rm c}} \sum_{{\bf k}}
\delta\bigl[ \mu - \tilde{\Sigma}_0(0) - E({\bf k}) 
\bigr] . 
\end{equation}
Physical properties of the Anderson model can be described
by the FL relation.\cite{yosida-yamada} 
%
%
In general, 
$2\tilde{\phi}_\gamma =\tilde{\phi}_s+\tilde{\phi}_c$, so that $1\le \tilde{\phi}_s/\tilde{\phi}_\gamma< 2$;
$\tilde{\phi}_s/\tilde{\phi}_\gamma$ is simply the Wilson ratio. \cite{wilsonKG,yosida-yamada}
When $n\rightarrow 1$ and $U/W\rightarrow+\infty$, in particular, 
charge fluctuations are totally suppressed, so that
$\tilde{\phi}_c/\tilde{\phi}_\gamma \rightarrow 0$ and
$\tilde{\phi}_s/\tilde{\phi}_\gamma \rightarrow 2$. In this paper, the Kondo temperature is defined by 
\begin{equation}\label{EqDefTK}
k_{\rm B}T_{\rm K} = 
1\big/\bigl[4\tilde{\phi}_\gamma \rho(0) \bigr] .
\end{equation}
The specific heat coefficient is given by
\begin{equation}\label{EqFLR-G}
\gamma = \frac{2}{3} \pi^2 k_{\rm B}^2 \tilde{\phi}_\gamma \rho(0)
=\frac{\pi^2 k_{\rm B}}{6T_{\rm K}} . 
\end{equation}

Physical properties of the Hubbard model can also be described
by the FL relation. \cite{Luttinger1,Luttinger2} 
%
According to the FS sum rule, 
\begin{equation}\label{EqFS-SumRule}
n = \frac{2}{N_{\rm c}} \sum_{{\bf k}}
\theta\bigl( [ \mu - \tilde{\Sigma}_0(0) - E({\bf k})
]/W \bigr) .
\end{equation}
According to Eqs.~(\ref{EqN0}) and (\ref{EqFS-SumRule}), 
\begin{equation}\label{EqMu-0}
\mu -\tilde{\Sigma}_0(0) = \mu_0(n).
\end{equation}
According to Eqs.~(\ref{EqRho0}), (\ref{EqRhoSSA1}) and (\ref{EqMu-0}), 
\begin{equation}\label{EqRhoSSA2}
\rho(0) =D \left[\mu_0(n)\right]. 
\end{equation}
According to Eqs.~(\ref{EqMappingCondition2}), and (\ref{EqExpandSelf}), 
\begin{equation}\label{EqImL-D}
\Delta(0) = 
\frac{\pi\rho(0)}{\left[\mbox{Re}R_\sigma(+i0)\right]^2 + 
\left[\pi\rho(0) \right]^2} .
\end{equation}
It should be noted that neither of $\rho(0)$ and $\Delta(0)$ depends on $U$ when $n$ is kept constant. 
The specific heat coefficient is given by Eq.~(\ref{EqFLR-G}). 
The Kondo temperature $T_{\rm K}$, which defined by Eq.~(\ref{EqDefTK}), is an energy scale of the effective Fermi energy of quasi-particles.

In the strong coupling regime defined by $U\agt W$, DOS has the three-peak structure, with the Gutzwiller band between UHB and LHB.
The Green function (\ref{EqGreen}) is approximately given at $T=0$~K by
\begin{equation}\label{EqGreen0}
G_\sigma(i\varepsilon_n, {\bf k}) =
\frac1{\tilde{\phi}_\gamma}
\frac1{i\varepsilon_n - \xi_0({\bf k})} + [\mbox{incoherent term}].
\end{equation}
The first term is the coherent term, which describes the Gutzwiller band. 
The dispersion relation and the band-width are given by
\begin{equation}\label{EqDisperSSA}
\xi_0({\bf k})= 
\bigl[E({\bf k}) +\tilde{\Sigma}_0(0) - \mu \bigr]/\tilde{\phi}_\gamma,
\end{equation}
and
\begin{equation}
W^*=W/\tilde{\phi}_\gamma \simeq 4k_{\rm B}T_{\rm K}.
\end{equation}
The incoherent term describes UHB and LHB.

\section{RVB mechanism and ordered states beyond S$^3$A or DMFT}
\label{SecBeyondSSA}
\subsection{Kondo-lattice theory}
\label{SecKL-theory}
The strong-coupling regime is mainly studied in this section.
The irreducible spin polarization function $\pi_s(i\omega_l,{\bf q})$ is also divided into single-site $\tilde{\pi}_s(i\omega_l)$ and multi-site $\Delta\pi_s(i\omega_l,{\bf q})$:
\begin{equation}\label{EqPi-1}
\pi_s(i\omega_l,{\bf q}) =
\tilde{\pi}_s(i\omega_l) +\Delta\pi_s(i\omega_l,{\bf q}) .
\end{equation}
The single-site $\tilde{\pi}_s(i\omega_l)$ is given by that of the Anderson model.
The spin susceptibilities of the Anderson and Hubbard
models are given, respectively, by
\begin{equation}\label{EqChi-1}
\tilde{\chi}_s(i\omega_l) =
\frac{2\tilde{\pi}_s(i\omega_l) }{
1 - U \tilde{\pi}_s(i\omega_l) }, 
\end{equation}
and
\begin{equation}\label{EqChi-2}
\chi_s(i\omega_l,{\bf q}) =
\frac{2\pi_s(i\omega_l,{\bf q}) }{
1 - U \pi_s(i\omega_l,{\bf q}) }.
\end{equation}
A physical picture for Kondo lattices is that local spin fluctuations on different sites interact by an intersite exchange interaction. According to this picture, an intersite exchange interaction $I_s(i\omega_l,{\bf q})$ is defined by
\begin{equation}\label{EqKondoKai}
\chi_s(i\omega_l,{\bf q}) =
\frac{\tilde{\chi}_s(i\omega_l) }{
1 - \frac1{4} I_s(i\omega_l,{\bf q}) \tilde{\chi}_s(i\omega_l) }.
\end{equation}
It follows from Eqs.~(\ref{EqPi-1}), (\ref{EqChi-1}), (\ref{EqChi-2}), and (\ref{EqKondoKai}) that
\begin{equation}\label{EqExch-I}
I_s(i\omega_l,{\bf q}) = 2 U^2 \Delta\pi_s(i\omega_l,{\bf q})
\left\{ 1 \hskip-1pt + \hskip-1pt
O\left[1/U\tilde{\chi}_s(i\omega_l)\right]\right\} .
\end{equation}
When $U/W\agt 1$, terms of $O[1/U\tilde{\chi}_s(i\omega_l)]$ can be ignored. 

The exchange interaction $I_s(i\omega_l,{\bf q}) $ is composed of three 
terms: \cite{three-exchange,itinerant-ferro}
\begin{equation}\label{EqThreeExchange}
I_s(i\omega_l,{\bf q}) =
J_s({\bf q}) + J_Q(i\omega_l,{\bf q})
- 4 \Lambda (i\omega_l,{\bf q}).
\end{equation}
The first term $J_s({\bf q})$ is the superexchange interaction:
\begin{equation}\label{EqJ-superNN}
J_s({\bf q}) = \frac1{N_{\rm c}}\sum_{\left<ij\right>}
Je^{i{\bf k}\cdot({\bf R}_i-{\bf R}_j) }.
\end{equation}
According to the second-order perturbation in $t$, where the widths of UHB and LHB are ignored, it follows that 
$J = - 4t^2/U$. \cite{andersonJ} 
According to field theory, the superexchange interaction arises from the virtual exchange of a pair excitation of electrons between UHB and LHB. \cite{three-exchange,itinerant-ferro,sup-exchange} 
When the widths of UHB and LHB are considered, 
$|J|$ is about a half of $4t^2/U$ when $U\simeq W$.\cite{exchange-reduction} 
The second term $J_Q(i\omega_l,{\bf q})$ 
arises from the virtual exchange of a pair excitation of quasi-particles.
When the single-site irreducible three-point vertex function in spin channels is denoted by $\tilde{\lambda}_s(i\varepsilon_n, i\varepsilon_n+i\omega_l; i\omega_l)$, it follows that
\begin{eqnarray}\label{Eq3pointVertex}
\tilde{\lambda}_s(0,0;0) &=& 
\tilde{\phi}_s[1 -U \tilde{\pi}_s(0)]
\nonumber \\ &= &
\frac{2 \tilde{\phi}_s}{U \tilde{\chi}_s(0) }
\left\{ 1 +O\left[1/U\tilde{\chi}_s(0) \right] \right\} ,
\end{eqnarray}
according to the Ward relation; \cite{ward} 
terms of $O[1/U\tilde{\chi}_s(0)]$ are also ignored. When only
the coherent part of the Green function is considered and 
Eq.~(\ref{Eq3pointVertex}) is approximately used for low-energy dynamical processes, 
\begin{equation}\label{EqJQ}
J_Q(i\omega_l,{\bf q}) = P(i\omega_l,{\bf q}) 
- \frac1{N_{\rm c}} \sum_{\bf q} P(i\omega_l,{\bf q}) ,
\end{equation}
with
\begin{eqnarray}\label{EqP-func}
P(i\omega_l,{\bm q}) &=&
\frac{4}{\tilde{\chi}_s^2(0)}
\left(\tilde{\phi}_s/\tilde{\phi}_\gamma \right)^2
\nonumber \\ && \times
\frac1{N_{\rm c}}\sum_{{\bf k}\sigma}
\frac{f[\xi_0({\bf k}) ]-f[\xi_0({\bf k} + {\bf q})]}
{i \omega_l - \xi_0({\bf k} + {\bf q}) + \xi_0({\bf k}) },
\qquad
\end{eqnarray}
which is derived in the random-phase approximation (RPA) for pair excitations of quasi-particles.
In Eq.~(\ref{EqJQ}), the single-site term is subtracted. 
The third term $- 4 \Lambda (i\omega_l,{\bf q})$ is the mode-mode coupling term among various types of fluctuations.
It corresponds to that in the self-consistent renormalization (SCR) theory of spin fluctuations, \cite{kawabata,moriya} which is relevant in the weak-coupling regime defined by $U/W\alt 1$. 

When Eq.~(\ref{Eq3pointVertex}) is approximately used, the mutual interaction mediated by spin fluctuations is given by
\begin{equation}\label{EqChi-J}
\frac1{4} \bigl[U \tilde{\lambda}_s(0,0;0)\bigr]^2
\bigl[\chi_s(i\omega_l,{\bf q}) - \tilde{\chi}_s(i\omega_l)\bigr]=
\frac1{4} \tilde{\phi}_s^2 I_s^*(i\omega_l,{\bf q}), 
\end{equation}
with 
\begin{equation}
I_s^*(i\omega_l,{\bf q})= \frac{ I_s(i\omega_l,{\bf q})}{
1 - \frac1{4} I_s(i\omega_l,{\bf q}) \tilde{\chi}_s(i\omega_l)}.
\end{equation}
In Eq.~(\ref{EqChi-J}), the single-site term is subtracted and two $\tilde{\phi}_s$ appear as effective
three-point vertex functions. The mutual
interaction mediated by spin fluctuations is simply the exchange interaction $I_s^*(i\omega_l,{\bf q})$. Multi-site or intersite terms are perturbatively considered in terms of $I_s(i\omega_l,{\bf q})$ or $I_s^*(i\omega_l,{\bf q})$. The perturbative theory is simply the Kondo-lattice theory.

\subsection{Stabilization of the FL by an RVB mechanism}
\label{SecFock}
There are two linear terms in the superexchange interaction $J_s({\bf q})$, which is the main term of $I_s^*(i\omega_l,{\bf q})$: Hartree-type\cite{mag-structure} and Fock-type \cite{phase-diagram} terms. The Hartree-type term, which gives magnetic Weiss mean fields, vanishes if no magnetic order parameter exists; magnetic instabilities are studied in Sec.~\ref{SecInstabilityFL}. When only the coherent term of the Green function is considered, the Fock-type term is given by
\begin{equation}\label{EqSelf-1}
\Delta \Sigma_\sigma ({\bf k}) = 
\frac{3}{4} \frac{\tilde{\phi}_{\rm s}^2}{\tilde{\phi}_\gamma}
\frac{k_{\rm B} T}{N_{\rm c}} 
\sum_{n{\bf p}}
J_s({\bf k}- {\bf p}) 
\frac{ e^{i \varepsilon_{n}0^+} }
{i\varepsilon_{n}- \xi_0({\bf p})} .
\end{equation}
The factor 3 appears because of three spin channels.
The Fock-type term $\Delta \Sigma_\sigma ({\bf k})$ should be self-consistently calculated with the single-site $\tilde{\Sigma}_\sigma(i\varepsilon_n)$ to satisfy the mapping condition (\ref{EqMappingCondition2}).
The self-consistent $\tilde{\Sigma}_\sigma(i\varepsilon_n)$ is expanded, as it is in Eq.~(\ref{EqExpandSelf}), but with renormalized $\tilde{\Sigma}_0(0)$, $\tilde{\phi}_\gamma$, $\tilde{\phi}_s$, and $\tilde{\phi}_c$ by the Fock-type term, all of which are real and finite. \cite{comMappingBeyondSSA} 
Then, the dispersion relation of quasi-particles is given by
\begin{equation}\label{EqDisperBeyond}
\xi({\bf k})= 
\bigl[E({\bf k}) +\tilde{\Sigma}_0(0) 
+ \Delta\Sigma_\sigma({\bf k}) - \mu \bigr]/\tilde{\phi}_\gamma,
\end{equation}
DOS at $\mu$ is given by
\begin{equation}\label{EqRhoBeyondSSA}
\rho(0) = \frac1{N_{\rm c}}\sum_{\bf k} \delta 
\bigl[E({\bf k}) +\tilde{\Sigma}_0(0) 
+ \Delta\Sigma_\sigma({\bf k}) - \mu \bigr] ,
\end{equation}
and the specific heat coefficient $\gamma$ is given by Eq.~(\ref{EqFLR-G}).

Consider the two dimensional square lattice $(d=2)$. 
The superexchange interaction is given by
%
$J_s({\bf q}) = 2J \left[\cos\left(q_xa\right)+\cos\left(q_ya\right) \right]$, 
with $a$ the lattice constant. It follows that
\begin{equation}\label{EqDeltaSigma}
\Delta \Sigma_\sigma ({\bf k}) =
\frac{1}{4}\tilde{\phi}_\gamma c_J J \left[\cos(k_xa) +\cos(k_ya)\right] , 
\end{equation}
with
\begin{equation}\label{EqXi}
c_J =
3 \left(\frac{\tilde{\phi}_{\rm s}}{\tilde{\phi}_\gamma}\right)^2
\hskip-3pt
\frac1{N_{\rm c}}\sum_{{\bf k}} \theta \left[\frac{-\xi_0({\bf k})}{W}\right]
%
\! \left[\cos(k_xa) +\cos(k_ya)\right] .
\end{equation}
Since $J$ is antiferromagnetic, the sign of $c_J$, which is $O(1)$, is such that $\Delta \Sigma_\sigma ({\bf k})$ enhances the band-width of quasi-particles.
The Fock-type term depends on $d$ and 
lattice structure.
In general, DOS is given by
\begin{equation}\label{EqRhoRed1}
\rho(0) \simeq 
1/\bigl[ W + \tilde{\phi}_\gamma |c_{J}J| \bigr],
\end{equation}
and the band-width of quasi-particles is given by 
\begin{equation}\label{EqW*}
W^* \simeq 1/\bigl[\tilde{\phi}_\gamma\rho(0)\bigr] 
\simeq W/\tilde{\phi}_\gamma + |c_{J} J| .
\end{equation}
The Kondo temperature is given by $k_{\rm B}T_{\rm K} \simeq W^*/4$.

Under S$^3$A or DMFT, the FL is stabilized by the Kondo effect, which stabilizes a local singlet on each unit cell.
The band-width of quasi-particles is $W/\tilde{\phi}_\gamma$, which may be infinitesimally small but is nonzero. Beyond S$^3$A or DMFT, the FL is further stabilized by the Fock-type term, which stabilizes a local singlet on each pair of nearest neighbors; the band-width is broadened by $|c_{J} J|$.
In this paper, the stabilization mechanism is called an RVB mechanism, since it is the same as or at least similar to that in the RVB theory. \cite{Plain-vanilla} 
If $\tilde{\phi}_\gamma \rightarrow +\infty$, in particular, the FL is totally stabilized by the RVB mechanism so that $W^*\simeq |c_J J|$.
Since DOS is vanishing such that $W\hskip-1pt\rho(0) \rightarrow 0$, the ground state is almost a spin liquid, which is studied in Appendix~\ref{SecSpin}, but is still the FL, provided that $U/W$ is finite.

It is an issue whether the FL is stable within the restricted Hilbert subspace where no order parameter exists.
In addition to the Fock-type term,
a few terms for the multi-site self-energy, \mbox{$\Delta\Sigma_\sigma(\varepsilon\!+\!i0,{\bf k})$}, 
are examined under an assumption that $\tilde{\phi}_\gamma<+\infty$ and no order parameter appears. 
In one dimension, terms proportional to $\varepsilon \ln |\varepsilon|$ appear at $T=0$~K, \cite{com1D} which
means that the FL is unstable against, al least, an exotic metal or the Tomonaga-Luttinger liquid. \cite{tomonaga,luttinger-L,kolomeisky} 
In $d\ge 2$ dimensions, no such term appear, which implies that the FL is stable, but within the restricted Hilbert subspace.

\subsection{Possible ordered states}
\label{SecInstabilityFL}
Since low-energy excitations are so accumulated in the FL that the specific heat is proportional to $T$ at \mbox{$T\ll T_{\rm K}$}, 
it is presumably the truth that, in the whole Hilbert space, the FL is eventually unstable against an ordered state in $d\ge 2$ dimensions.
When an order parameter is specified, it is straightforward to study the response function corresponding to it, from which the instability condition of the FL against the ordered state can be derived. 
In this paper, the phase diagram is out of scope but only possible ordered states are examined.

The magnetic susceptibility $\chi_s(i \omega_l,{\bf q})$ is given by Eq.~(\ref{EqKondoKai}).
Provided that 
\begin{equation}
\bigl[1-\mbox{$\frac1{4}$}I_s(0,{\bf Q})\tilde{\chi}_s(0)\bigr]_{T=0\hskip1pt{\rm K}}>0 ,
\end{equation}
for any ${\bf q}$, the FL is stable against any magnetic state.
Assume that $I_s(0,{\bf q})$ is maximal at ${\bf q}={\bf Q}$.
When \mbox{$I_s(0,{\bf Q})$ or $\tilde{\chi}_s(0)$} is so large that
\begin{equation}
\left[1 - \mbox{$\frac1{4}$} I_s(0,{\bf Q})\tilde{\chi}_s(0)\right]_{T=T_{\rm N}} = 0,
\end{equation}
the FL is unstable below $T_{\rm N}$ against 
a magnetic state or a spin density wave (SDW) state with ${\bf Q}$.
According to the FL relation, \cite{yosida-yamada} together with Eq.~(\ref{EqDefTK}), it follows that $\tilde{\chi}_s(0)\simeq 1/k_{\rm B} T_{\rm K}$. In the limit of $n\rightarrow 1$ and $U/W\rightarrow +\infty$, 
$k_{\rm B}T_{\rm K}$ is vanishing so that $\tilde{\chi}_s(0)$ is diverging.
The superexchange interaction $J_s({\bf q})$ is antiferromagnetic. 
When $n\simeq 1$, $J_Q(0,{\bf q})$ is also antiferromagnetic, as is discussed in Sec.~\ref{SecFerro}.
When $n\simeq 1$ and $U/W$ is large enough, therefore, it is probable that the FL is unstable against an antiferromagnetic state.
In two dimensions, however, $T_{\rm N}=+0$~K 
when the mode-mode coupling term, $- 4 \Lambda (0,{\bf q})$, or magnetic critical fluctuations are self-consistently treated. \cite{mermin}
 

An average of $I_s^*(\omega+i0,{\bf q})$ 
over a low-energy region $|\omega|\alt k_{\rm B}T_{\rm K}$ is expanded as 
\begin{equation}
\mbox{Re} \bigl<I_s^*(\omega + 
i0,{\bf q})\bigr>_{\omega} \hskip-1pt =
\frac1{N_{\rm c}}
\sum_{ij} I_{ij}^* 
e^{i {\bf q}\cdot ({\bf R}_i - {\bf R}_j)},
\end{equation}
with $\left<\cdots\right>_{\omega}$ standing for the average.
Most possible order parameters, in addition to a magnetic one, are given by the decoupling of 
\begin{equation}
 - \frac1{2} \sum_{i\ne j}
\sum_{\alpha\beta\gamma\delta} I_{ij}^* 
\left(\mbox{$\frac1{2}$}{\bm \sigma}^{\alpha\beta}\right)\cdot
\left(\mbox{$\frac1{2}$}{\bm \sigma}^{\gamma\delta}\right)
a_{i\alpha}^\dagger a_{i\beta} a_{j\gamma}^\dagger a_{j\delta },
\end{equation}
with ${\bm \sigma}=(\sigma_x,\sigma_y,\sigma_z)$ being the Pauli matrix:
$\bigl< a_{i\tau}^\dagger a_{j\tau^\prime }^\dagger\bigr> $ or $\bigl< a_{j\tau^\prime} a_{i\tau} \bigr>$ of superconductivity, 
$\sum_{\tau} \bigl< a_{i\tau}^\dagger a_{j\tau}\bigr> $ of charge bond wave (CBW), and
$\sum_{\tau\tau^\prime} \sigma_{\nu}^{\tau\tau^\prime}\bigl< a_{i\tau}^\dagger a_{j\tau^\prime }\bigr> $ of spin bond wave (SBW), in addition to $\sum_{\tau\tau^\prime} \sigma_{\nu}^{\tau\tau^\prime} \bigl< a_{i\tau}^\dagger a_{i\tau^\prime }\bigr>$ of magnetism or SDW.

For the sake of simplicity, the nearest-neighbor component of $I_{ij}^*$, which is denoted by $I^*_1$, is only considered.
When $n\simeq 1$, $J_s({\bf q})$ and $J_Q(0,{\bf q})$ are antiferromagnetic, as is discussed above, so that $I_1^*$ is antiferromagnetic such that $I_1^*<0$. 
When $n\simeq 0$ or $n\simeq 2$, $J_Q(0,{\bf q})$ can be ferromagnetic, as is discussed in Sec.~\ref{SecFerro}. 
When $J_s({\bf q})$ is weak, $I_1^*$ can be ferromagnetic such that $I_1^*>0$. Possible symmetries or waves of ordered states depend on $d$, lattice structure, $n$, the sign of $I_1^*$, and others.

When $I_1^*$ is weak or strong, the FL is unstable against an anisotropic superconducting (SC) state, at least, if no disorder exists:
a singlet one for $I_1^*<0$ and a triplet one for $I_1^*>0$. 
For example, consider the two-dimensional square lattice. When $n\simeq 1$, $I_1^*<0$, as is discussed above. Then, two singlet waves are possible: anisotropic $s$ wave and $d\gamma$ wave.
When any pair breaking by SC critical fluctuations themselves and other intersite fluctuations are ignored, \cite{com-Onsite} SC critical temperatures $T_c$ are given by
\begin{equation}\label{EqSC-Tc}
1 + \frac{3}{4}I_1^* 
\left(\frac{\tilde{\phi}_s}{\tilde{\phi}_\gamma}\right)^2
\hskip-3pt \frac1{N_{\rm c}}\sum_{\bf k}
\frac{\eta_{\Gamma}^2({\bf k})}{\xi({\bf k})}\tanh
\hskip-1pt
\left[\frac{\xi({\bf k})}{2k_{\rm B}T_{\rm c}}\right] =0,
\end{equation}
with 
\begin{equation}
\eta_{\Gamma}({\bf k}) = \left\{\begin{array}{ll}
\cos(k_xa)+\cos(k_ya), & \mbox{ $\Gamma=s$}  \\
\cos(k_xa)-\cos(k_ya), & \mbox{ $\Gamma=d\gamma$} 
\end{array}\right. ,
\end{equation}
being form factors of the $s$ wave and the $d\gamma$ wave. 
Equation~(\ref{EqSC-Tc}) gives approximate or at least upper limits of $T_c$'s in quasi-two dimensions; $T_c=+0$~K in two dimensions because of SC critical fluctuations. \cite{mermin} 
In quasi-two dimensions, when $n\simeq 1$, $T_c$ of the $d\gamma$ wave is much higher than $T_c$ of the $s$ wave since $|\eta_{d\gamma}({\bf k})|^2 \gg |\eta_{s}({\bf k})|^2$ for almost ${\bf k}$'s except for $|k_x|\simeq |k_y|$ on the FS. \cite{com-Onsite}
When $I_1^*<0$, triplet $p$ waves are possible: $p_x$ and $p_y$ waves.

The homogeneous CBW is simply the Fock-type term studied in Sec.~\ref{SecFock}. Since it breaks no symmetry at least in the FL, it is not an ordered parameter. 
When $I_1^*$ is weak, the FL is stable against any bond wave. When $I_1^*$ is strong, the FL is unstable against CBW for $I_1^*<0$ and SBW for $I_1^*>0$. 
A flux state, which is a multi-${\bf Q}$ bond wave with different phases for different ${\bf Q}$ components, is also possible, with ${\bf Q}$ ordering wave numbers. \cite{lee}


When \mbox{$U/W\alt 1$}, the perturbation in $U$ is more 
useful than that in $I_s(i\omega_l,{\bf q})$ is. When the nesting of the FS is sharp enough, the FL is unstable against an SDW state. When $U^2\chi_s(i\omega_l,{\bf q})$ is considered as an interaction between electrons, it is unstable against an anisotropic SC state, \cite{kondo} at least, if no disorder exists.

The above analysis can never exclude a possibility of an exotic state. If it is characterized by an order parameter, it is straightforward to study the instability of the FL against it.

\section{Discussion}
\label{SecDiscussion}
\subsection{$1/d$ expansion theory} 
\label{Sec1dExpansion}
Every term or quantity is classified according to the order in $1/d$ in the site \cite{Metzner} and wave-number \cite{Mapping-2,Mapping-3} representations. The single-site $\tilde{\Sigma}_\sigma(i\varepsilon)$ is of leading order in $1/d$ or $O(1)$. When $U/|t|>0$, \cite{comAttractive} the single-site $\tilde{\chi}_s(i\omega_l)$ is a relevant $O(1)$ term. Multi-site or intersite terms can also be $O(1)$ only for particular ${\bf Q}$'s in the wave-number representation. Relevant $O(1)$ terms are the magnetic $J_s({\bf Q})$ and $J_Q(i\omega_l, {\bf Q})$ for the particular ${\bf Q}$'s. On the other hand, 
$\Delta\Sigma_\sigma(i\varepsilon_n,{\bf k})=O(1/\sqrt{d})$ for any ${\bf k}$, $J_s({\bf q})=O(1/\sqrt{d})$ and $J_Q(i\omega_l, {\bf q})=O(1/\sqrt{d})$ for almost all ${\bf q}$'s except for the particular ${\bf Q}$'s, and $\Lambda (i\omega_l,{\bf q})=O(1/d)$ for any ${\bf q}$. 

When the N\'{e}el temperature $T_{\rm N}$ is nonzero for one of the particular ${\bf Q}$'s, magnetization ${\bf m}({\bf Q})$ appears at $T<T_{\rm N}$. Magnetic Weiss mean fields, $J_s({\bf Q}){\bf m}({\bf Q})$ and $J_Q(0, {\bf Q}){\bf m}({\bf Q})$, are $O(1)$. When they are considered in the mean-field approximation beyond S$^3$A or DMFT, the mean-field theory is rigorous for $d=+\infty$. 
When $T\ge T_{\rm N}$, S$^3$A or DMFT is rigorous for $d=+\infty$, except for the magnetic susceptibility with the particular ${\bf Q}$'s. \cite{comAttractive}

%

The cluster DMFT (CDMFT) is a non-perturbative theory for a cluster to include multi-site terms beyond S$^3$A or DMFT. \cite{cellDMFT1,cellDMFT2,cellDMFT3,cellDMFT4}
In CDMFT, the translational symmetry is broken by choosing of a particular cluster even in the presence of no order parameter; 
the symmetry is recovered for an infinitely large cluster.
The Kondo-lattice theory is a perturbative theory to include multi-site terms starting from the FL in S$^3$A or DMFT, which is a non-perturbative theory. 
The conventional perturbation can treat higher-order terms in $1/d$; it can also treat instability of the FL, as is examined in Sec.~\ref{SecInstabilityFL}. The anomalous perturbation that assumes the existence of an order parameter can treat magnetic order, which is $O(1)$, and other types of order such as anisotropic superconductivity and bond wave, which are of higher order in $1/d$. \cite{comAttractive} 
In the Kondo-lattice theory, the translational symmetry is not broken by its framework itself; it is broken when an order parameter with nonzero wave-number appears. 

\subsection{Magnetism crossover} 
\label{SecMagnetismCrossover}
The study in this paper is almost restricted to $T=0$~K, except for the study in Sec.~\ref{SecInstabilityFL}. It is straightforward to extend the study to $T>0$~K. 
In the Kondo effect, the $T$ dependent crossover occurs between a local-moment magnet at $T\gg T_{\rm K}$ and the FL at $T\ll T_{\rm K}$.\cite{wilsonKG} In a magnet, there exists a temperature scale $T_{\rm N}^*$, below which magnetic critical fluctuations develop. 
When $T_{\rm N}^* \gg T_{\rm K}$, the magnet is characterized as a typical local-moment one. When $T_{\rm N}^* \ll T_{\rm K}$, it is characterized as a typical itinerant-electron one.
The magnetism crossover is simply a crossover between local-moment magnetism at $T\agt T_{\rm K}$ and itinerant-electron magnetism at $T\alt T_{\rm K}$.

The local susceptibility $\tilde{\chi}_s(0)$ gives the Curie-Weiss (CW) law for any ${\bf q}$ or the CW law of a local-moment magnet at $T\agt T_{\rm K}$.\cite{wilsonKG} The RPA polarization function $J_Q(0,{\bf q})$ or $P(0,{\bf q})$ gives the CW law of an itinerant electron magnet at $T\alt T_{\rm K}$. When there is a sharp nesting of FS, $P(0,{\bf q})$ gives the CW law only around the nesting wave number.\cite{fjo-CW} When the chemical potential lies around a sharp peak of DOS, $P(0,{\bf q})$ gives the CW law only for ${\bf q}\simeq 0$.\cite{miyai} The two mechanisms are $O(1)$ in $1/d$. The particular ${\bf q}$ dependences of the two CW mechanisms are remnants of those in infinite dimensions. 

When there is no nesting of FS and DOS is almost constant around the chemical potential, the mode-mode coupling term $- 4 \Lambda (0,{\bf q})$ can give a local-moment type CW law of an itinerant electron liquid at $T\alt T_{\rm K}$.
\cite{moriya}
When there is a sharp nesting of FS or the chemical potential is around a sharp peak of DOS, the mode-mode coupling term shows an opposite temperature dependence to the CW law.\cite{miyake,miyai}
Since $- 4 \Lambda (0,{\bf q})=O(1/d)$, this mechanism does not work in infinite dimensions.

The magnetism crossover in the periodic Anderson model (PAM) is slightly different from that in the Hubbard model. In PAM, conduction electrons and the so called $d$ or $f$ electrons are strongly hybridized at $T\alt T_{\rm K}$ to form heavy electrons or quasi-particles, while they are independent degrees of freedom at $T\gg T_{\rm K}$ such that $d$ or $f$ electrons behave as localized spins but conduction electrons are itinerant. 
At $T\alt T_{\rm K}$, there is no essential difference between the two models. The exchange interaction $J_Q(i\omega_l,{\bf q})$, which arises from the virtual exchange of a pair excitation of quasi-particles, works between quasi-particles themselves in either model. At $T\gg T_{\rm K}$, there is a crucial difference between the two models. In PAM, the Ruderman-Kittel-Kasuya-Yosida (RKKY) exchange interaction $J_{\rm RKKY}(i\omega_l,{\bf q})$ arises from the virtual exchange of a pair excitation of conduction electrons, and it works between localized spins of $d$ or $f$ electrons in PAM.
The $d$ or $f$ electron component of the susceptibility of PAM is given by Eq.~(\ref{EqKondoKai}) with
\begin{equation}
I_s(i\omega_l,{\bf q}) = J_s({\bf q}) + J_{\rm RKKY}(i\omega_l,{\bf q})
- 4 \Lambda (i\omega_l,{\bf q}),
\end{equation}
at $T\gg T_{\rm K}$; the superexchange interaction $J_s({\bf q}) $ also exists in PAM, in general. When $T\alt T_{\rm K}$, $J_{\rm RKKY}(i\omega_l,{\bf q})$ turns out to be $J_{Q}(i\omega_l,{\bf q})$. \cite{three-exchange}
The competition between the Kondo effect and the RKKY exchange interaction is only relevant at $T\gg T_{\rm K}$ or $T\agt T_{\rm K}$. 

\subsection{Evidence for the existence of $J_Q(i\omega_l,{\bf q})$}
\label{SecFerro}
The exchange interaction $J_Q(i\omega_l,{\bf q})$ or $P(i\omega_l,{\bf q})$ is responsible for magnetic properties that are observed in itinerant-electron magnets.
It is similar to the conventional RPA polarization function except for the pre-factor of $\bigl[4/\tilde{\chi}_s^2(0)\bigr]
\bigl(\tilde{\phi}_s/\tilde{\phi}_\gamma \bigr)^2$.
Because of this factor, its strength is proportional to $W^*\simeq 4k_{\rm B}T_{\rm K}$.\cite{satoh1,satoh2} \ 
In particular, 
\begin{equation}
\lim_{|{\bf q}|\rightarrow 0} P(0,{\bf q})=
\frac{8}{\tilde{\chi}_s^2(0)}
\left(\tilde{\phi}_s^2/\tilde{\phi}_\gamma \right)\rho(0) \propto k_{\rm B}T_{\rm K},
\end{equation}
for the static ${\bf q}=0$ component. Here, $\tilde{\chi}_s(0) \simeq 1/(k_{\rm B}T_{\rm K})$ is used.
Provided that $|J_Q(0,{\bf q})| \gg |J_s({\bf q})- 4 \Lambda (0,{\bf q})|$, low-energy phenomena are characterized by a single energy scale of $k_{\rm B}T_{\rm K}$, i.e., physical properties obeys the so called one-parameter scaling. 
The one-parameter scaling in $k_{\rm B}T_{\rm K}$ is actually observed in the metamagnetic transition or crossover in CeRu$_2$Si$_2$,
\cite{haen1,haen2,haen3,sakakibara}
which is evidence that 
$J_Q(i\omega_l,{\bf q})$ is relevant at least in CeRu$_2$Si$_2$.

It is obvious that $P(0,{\bf q})$ has similar features to those of the conventional RPA polarization function.
For example, it is antiferromagnetic when the nesting of FS of quasi-particles is sharp or the chemical potential lies around the center of the quasi-particle band or $n\simeq 1$. It is ferromagnetic when the chemical potential lies around the top or bottom of the quasi-particle band, i.e., for $n\simeq 2$ or $n\simeq 0$. In particular, it is strongly ferromagnetic when DOS has a sharp peak at one of the band edges where the chemical potential lies,\cite{itinerant-ferro,satoh1,satoh2}
as DOS's of many itinerant-electron ferromagnets such as Fe, Ni and so on have. This is consistent with that of Kanamori's theory for itinerant-electron ferromagnetism.\cite{kanamori} 
Since the superexchange interaction is antiferromagnetic, however, it seems to be difficult for itinerant-electron ferromagnetism to occur in the {\it single-band} Hubbard model.

According to Ref.~\onlinecite{itinerant-ferro}, 
the superexchange interaction is ferromagnetic in a {\it multi-band} Hubbard model if the Hund coupling is strong enough and the band degeneracy is large enough. Itinerant-electron ferromagnetism can easily occur when both of the superexchange interaction and $J_Q(0,{\bf q})$ are ferromagnetic.

\subsection{Impossibility of the Mott insulator for finite $U$}
Since the proof in Sec.~\ref{SecProofGround} is made for $\lambda=\pm0$, it is another issue what is the ground state in {\it the grand canonical ensemble with $\lambda=0$} or the canonical ensemble, where ${\cal N}$ or $N$ is a conserved quantity. When $\lambda=0$, the FS condition may be or may not be satisfied so that the ground state of the Anderson model may be or may not be a singlet.
If it is not a singlet, the ground state of the Hubbard model under S$^3$A or DMFT is an insulator whose entropy is diverging in the thermodynamic limit. 
In this paper, only such an abnormal insulator is called the Mott insulator; if the ground-state entropy of an insulator with $n\simeq1$ or $N\simeq N_{\rm c}$ is zero or finite in the thermodynamic limit, the insulator is called a spin liquid.
Since it is quite unlikely that the third law of thermodynamics is broken in a relevant Hamiltonian such as the Hubbard model with finite $U/|t|$, we speculate that the ground state under S$^3$A or DMFT is a singlet for even $N$ or a doublet for odd $N$.
Even if the ground state can be infinitely degenerate under S$^3$A or DMFT, the degeneracy must be lifted when the RVB mechanism is considered beyond it.
We also speculate that when $U/|t|$ is finite the ground state within the restricted Hilbert subspace must be the FL, an exotic metal, or a spin liquid for any $d$ in either case of $\lambda=0$ and $\lambda=\pm 0$.

According to Lieb and Wu's Bethe-ansatz solution for one dimension in the canonical ensemble, \cite{Lieb-Wu} when $N=N_{\rm c}$, an \mbox{M-I} transition occurs at $U=0$. The singularity at $U=0$ is exotic. \cite{Takahashi1} The \mbox{M-I} transition at $U=0$ is never due to the opening of the Hubbard gap.
The insulator for finite $U/|t|$ is Lieb and Wu's insulator or spin liquid rather then the Mott insulator. \cite{Takahashi2} When $N\ne N_{\rm c}$ and $U/|t|>0$, the ground state is the Tomonaga-Luttinger liquid, \cite{tomonaga,luttinger-L,kolomeisky} in which the charge-spin separation occurs. \cite{comSeparation} 
When $U/|t|= +\infty$, in particular, the liquid is an abnormal metal whose ground-state entropy is diverging in the thermodynamic limit. Because of the complete exclusion of double occupancy and only the transfer integral $t$ between nearest neighbors being nonzero, the charge-spin separation is complete such that any eigen-function is a Cartesian product of charge and spin parts, i.e., a Slater determinant of spinless fermions and an eigen-function of non-interacting localized $S=1/2$ spins.\cite{kawakami,ogata} The ground-state entropy is $Nk_{\rm B}\ln 2$ and the spin susceptibility obeys the Curie law. The charge susceptibility is nonzero and it diverges as $n\rightarrow 1-0$ and $n\rightarrow+0$ at $T=0$~K, with $n=N/N_{\rm c}$. When $N\ne N_{\rm c}$ and $U/|t|=+\infty$, {\it spins} are localized but {\it charges} are itinerant. 
When $N\ne N_{\rm c}$, the ground state is an exotic or abnormal metal.

In general, electrons are more itinerant in $d\ge 2$ dimensions than they are in one dimension. It is likely that the ground state within the restricted Hilbert subspace is a metal at least for $n\ne 1$ or $N\ne N_{\rm c}$. In general, the nature of electron correlations is less abnormal in $d\ge 2$ dimensions than it is in one dimension. It is likely that, provided that $U/|t|$ is finite, the ground state within the restricted Hilbert subspace is a singlet in the grand canonical ensemble and is a singlet or a doublet in the canonical ensemble.

When the Mott \mbox{M-I} transition is studied in the Gutzwiller approximation, the canonical ensemble is conventionally assumed. 
The abnormal insulator for $N=N_{\rm c}$ and $U\ge U_{\rm BR}$ according to Brinkman and Rice's theory \cite{brinkman} must be unstable when the RVB mechanism is considered. According to the proof in Sec.~\ref{SecProofGround}, any insulator is unstable for finite $U/|t|$ in the grand canonical ensemble with $\lambda=\pm 0$ even if the RVB mechanism is not considered, i.e., under S$^3$A or DMFT, which is beyond the Gutzwiller approximation.

In numerical S$^3$A or DMFT, \cite{RevMod,PhyToday,kotliar} 
CDMFT,\cite{cellDMFT1,cellDMFT2,cellDMFT3,cellDMFT4}
and Monte Carlo theory, \cite{imada1,imada2,imada3} 
an \mbox{M-I} transition seems to occur when $n\simeq 1$ and $U\simeq U_{\rm BR}$.
The \mbox{M-I} transition, at least, for $n\ne 1$ seems to inconsistent with the analysis of this paper, although the phase diagram for $T>0$~K, which is studied in numerical theories, may be different from that for $T=0$~K, in general. 
First of all,
the effect of $\lambda=\pm0$ or $N$ being a non-conserved quantity in the grand canonical ensemble, if it is not considered, should be explicitly considered also in numerical theories because single-particle excitations in a system where $N$ is not a conserved quantity are different from those in a system where $N$ is a conserved quantity, as is studied in Appendix~\ref{SecAppendixGutzwiller}.
When an M-I transition occurs in either of the numerical theories, a lower-temperature phase seems to be the Mott insulator.\cite{PhyToday} 
It is interesting to examine which is actually observed in numerical theories, evidence that the third law of thermodynamics holds or evidence that it does not. In this context, it is interesting to carry out numerical processes beyond S$^3$A or DMFT in a parameter region where the RVB mechanism is expected to be effective. 
If a stabilization effect is observed, it is evidence that the ground state is a singlet under and beyond S$^3$A or DFFT.
If it is not observed, the ground state may be infinitely degenerate in the thermodynamic limit, but within the restricted Hilbert subspace; the true ground state must be an ordered state in the whole Hilbert space. If this is the truth, it should be clarified what impedes the RVB mechanism or what stabilizes the abnormal ground state, where the third law of thermodynamics is broken.

\subsection{Normal state of cuprate superconductors}
\label{SecSuperconductivity}
If the RVB state \cite{RVB,fazekas} is characterized by an order parameter, it is straightforward to study the instability of the FL against it. However, no order parameter is proposed so far.
In the mean-field RVB theory,\cite{Plain-vanilla}
the RVB state is stabilized by local correlations, which are treated with the Gutzwiller projection operator, and the Fock-type exchange interaction. 
No order parameter is introduced by the Gutzwiller projection operator, which is also used to treat the FL, or the Fock-type term, which is also nonzero in the FL. 
On the other hand, it is speculated that the charge-spin separation occurs in the RVB state; bosonic and fermionic elementary excitations are called holons and spinons. \cite{RVB} In order to support this speculation,
a slave-boson RVB theory is proposed in the {\it slave-boson} $t$-$J$ model; \cite{fukuyama1,fukuyama2} an electron in the $t$-$J$ model corresponds to a pair excitation of a slave boson and a fermion in the slave-boson $t$-$J$ model. In this theory, the RVB state is characterized by the condensation of slave bosons, which are called {\it holons}; itinerant fermions, which are called {\it spinons}, appear when slave bosons are condensed.
However, it is never shown so far what symmetry is broken in the \mbox{$t$-$J$} model when the condensation of slave bosons occurs in the slave-boson $t$-$J$ model. 
There is no evidence that the RVB state in the $t$-$J$ model is characterized by an order parameter, which implies that the RVB state, which is an insulator or a metal, is of the same symmetry as the FL is.
On the basis of the FL theory for the Heisenberg model in Appendix~\ref{SecSpin}, which implies that the adiabatic continuity \cite{AndersonText} holds between a spin liquid in the Heisenberg model and the FL in the Hubbard model, we propose that the insulating RVB state is the spin liquid and the metallic RVB state is the FL.

Under S$^3$A or DMFT, $\rho(0)\simeq 1/W$, as is shown in Eq.~(\ref{EqRhoSSA2}). Beyond S$^3$A or DMFT, $\rho(0)$ is reduced by the RVB mechanism such that
$\rho(0) \simeq 
1/\bigl[ W + \tilde{\phi}_\gamma |c_{J}J| \bigr]$, as is shown in Eq.~(\ref{EqRhoRed1}).
On the other hand, DOS's of LHB and UHB are not reduced by the RVB mechanism.
In the so called Hubbard III approximation, \cite{Hubbard2} the band-width of UHB and LHB is $W_{\rm HB}\simeq 0.8W$ for $n\simeq 1$.
DOS's of LHB and UHB are given by 
\begin{equation}
\rho(\epsilon_a-\mu)\simeq \rho(\epsilon_a+U-\mu)
\simeq 1/(2W_{\rm HB}).
\end{equation}
%
When $n\simeq 1$ and $n<1$, LHB is just below the quasi-particle band.
The band centers of LHB and UHB are $\epsilon_a-\mu\simeq -W_{\rm HB}/2$ and $\epsilon_a +U -\mu\simeq U -W_{\rm HB}/2$. 
LHB spreads over $- W_{\rm HB}\alt \varepsilon \alt -W^*/2$, the quasi-particle band over $|\varepsilon|\alt W^*/2$, and UHB over $U- W_{\rm HB}\alt \varepsilon \alt U$. 
When UHB is totally above the quasi-particle band, an energy region of $\varepsilon \simeq - W^*$ is within LHB and that of $\varepsilon \simeq W^*$ is within a gap-region between the quasi-particle band and UHB, so that
%
$\rho(\varepsilon\simeq -W^*)\gg \rho(\varepsilon\simeq W^*)$.
%
If the reduction of $\rho(0)$ is small or $\rho(0)\simeq 1/W$, 
\begin{subequations}\label{EqObsSym0}
\begin{equation}\label{EqObsSym2}
\rho\bigl(\varepsilon\simeq -W^*\bigr) \simeq \rho(0) 
\gg \rho(\varepsilon\simeq W^*). 
\end{equation} 
If the reduction of $\rho(0)$ is large or $\rho(0)\ll 1/W$, 
\begin{equation}\label{EqObsSym3}
\rho\bigl(\varepsilon\simeq -W^*\bigr) \gg \rho(0) 
\simeq \rho(\varepsilon\simeq W^*). 
\end{equation} 
\end{subequations}
An asymmetric $\rho(\varepsilon)$ that is consistent with Eq.~(\ref{EqObsSym0}) is observed by tunneling spectroscopy.\cite{asymmetry1}
Since such an asymmetry can arise from the reduction of $\rho(\varepsilon\simeq 0)$, it is evidence that the RVB mechanism is crucial in cuprate superconductors. 

In two dimensions, critical fluctuations make $T_{\rm c}$ down to $+0$~K. \cite{mermin} This implies that large deviations from the typical FL can occur in anisotropic quasi-two dimensions and they are responsible for some of exotic properties of cuprate superconductors. 
If the anisotropy is large enough but $T_{\rm c}$ is still high enough, SC critical fluctuations can cause the opening of a pseudo-gap in the SC critical region. \cite{FJO-PsGap1,FJO-PsGap2} 
It is certain that other effects, such as the electron-phonon interaction \cite{el-ph1,el-ph2} and so on, are necessary to explain the whole exotic properties. \cite{RevD-wave,RevScience,RevStripe,lee,RevScanning} 

Within the Hubbard model, $T_{\rm c}$ is low since $|J|$ is about a half of $4|t|^2/U$ in an actual situation. \cite{exchange-reduction} Experimentally, $J$ is as large as $J\simeq -0.15$~eV. \cite{SuperJ} In order to explain observed $T_{\rm c}$, the phenomenological $J\simeq -0.15$~eV should be used in Eq.~(\ref{EqJ-superNN}) or the $t$-$J$ model with $J\simeq -0.15$~eV should be used. The $d$-$p$ model with relevant parameters corresponds to the $t$-$J$ model. \cite{rice} If $T_{\rm c}$ should be explained microscopically, the $d$-$p$ model should be used. \cite{exchange-reduction} 
It is straightforward to extend the study of this paper to the $t$-$J$ model and the $d$-$p$ model.
It is also straightforward to extend the study of this paper further such that effects of SC, magnetic, and bond wave fluctuations can be included, not only in the Hubbard model but also in the $t$-$J$ model and the $d$-$p$ model.

According to an early FL theory of high-$T_{\rm c}$ super\-con\-duc\-tiv\-i\-ty in 1987, \cite{highTc1,highTc2} the normal state is the FL and the condensation of $d\gamma$-wave Cooper pairs is responsible for high-$T_{\rm c}$ superconductivity.
The analysis on possible SC states in this paper is simply an extension of the early FL theory, or it confirms the early FL theory although the extensions are necessary.

\section{Conclusion}
\label{SecConclusion}
The Hubbard model is studied by the Kondo-lattice theory.
The supreme single-site approximation (S$^3$A), which considers all the single-site terms, is reduced to determining and solving self-consistently the Anderson model, which is an effective Hamiltonian for the Kondo effect.
It is proved that the ground state under S$^3$A is the Fermi liquid except for $n=1$ and $U/W=+\infty$, with $n$ being the electron density per unit cell, $U$ the on-site repulsion, and $W$ the band-width.
Multi-site terms are perturbatively considered beyond S$^3$A by the Kondo-lattice theory.
When $n\simeq 1$ and $U/W\agt 1$, in particular, the Fermi liquid is stabilized under S$^3$A by the Kondo effect and is further stabilized beyond S$^3$A by the resonating valence bond (RVB) mechanism.  
The Fermi liquid is a relevant normal state to study possible lower-temperature phases or the true ground state. 
In one dimension, the Fermi liquid is unstable against, at least, the Tomonaga-Luttinger liquid. In two dimensions and higher, the Fermi liquid is unstable against, at least, an antiferromagnetic or anisotropic superconducting state.

It is proposed that the Fermi liquid stabilized by the Kondo effect and the RVB mechanism is the normal state of cuprate superconductors. 
In order to explain high superconducting critical temperatures, however, the $t$-$J$ model with $J\simeq -0.15~$eV or the $d$-$p$ model with relevant parameters, which corresponds to to the $t$-$J$ model, should be used instead of the Hubbard model.

\section*{Acknowledgements}
The author is thankful to M. Ido, M. Oda and N.
Momono for useful discussions on various properties of cuprate superconductors.

\appendix

\section{Proof of the Inequality~(\ref{EqImp})} 
\label{SecProof}
Define the following real functions:
\begin{equation}
S_1(\varepsilon,\varepsilon^\prime) =
\varepsilon + \mu - \varepsilon^{\prime}
- \mbox{Re} \bigl[ 
 \tilde{\Sigma}_\sigma(\varepsilon \!+\! i0)
-i \lambda^2 \Gamma (\varepsilon \!+\! i0)
\bigr],
\end{equation}
\begin{equation}\label{EqAppemdixS2}
S_2(\varepsilon)= 
-\mbox{Im} \bigl[ 
 \tilde{\Sigma}_\sigma(\varepsilon + i0)
-i \lambda^2 \Gamma (\varepsilon + i0) \bigr] ,
\end{equation}
and 
\begin{equation}\label{EqAppYN}
Y_n(\varepsilon) = \int d\varepsilon^\prime
D(\varepsilon^\prime) 
\frac{S_1^n(\varepsilon,\varepsilon^\prime) }
{S_1^2(\varepsilon,\varepsilon^\prime) +S_2^2(\varepsilon)}.
\end{equation}
In this Appendix, it is only assumed that $\tilde{\Sigma}_\sigma(\varepsilon + i0)$ is an analytical function in the upper half plane; it may be convergent or divergent on the real axis.
It follows from Eq.~(\ref{EqAppYN}) that
\begin{equation}\label{EqY2}
Y_2(\varepsilon) =1 - S_2^2(\varepsilon) Y_0(\varepsilon).
\end{equation}
Since $R_\sigma(\varepsilon+i0)$ defined by Eq.~(\ref{EqR-delta1}) is given by 
\begin{equation}
R_\sigma(\varepsilon+i0) =
Y_1(\varepsilon) - i S_2(\varepsilon) Y_0(\varepsilon) ,
\end{equation}
the mapping condition (\ref{EqMappingCondition2}) is given by 
\begin{equation}\label{EqAppendixL}
\Delta(\varepsilon) =
\mbox{Re}\lambda^2 \Gamma(\varepsilon \!+\! i0) 
\hskip-1pt - \hskip-1pt S_2(\varepsilon)
\frac{Y_1^2(\varepsilon) 
\hskip-1pt - \hskip-1pt Y_0(\varepsilon) Y_2(\varepsilon)}
 {Y_1^2(\varepsilon) \hskip-1pt + \hskip-1ptS_2^2(\varepsilon) 
Y_0^2(\varepsilon)} .
\end{equation}
Here, Eq.~(\ref{EqY2}) is used.
It is trivial that 
$S_2(\varepsilon) >0$
and 
$Y_1^2(\varepsilon) + S_2^2(\varepsilon) Y_0^2(\varepsilon) > 0$.
Since an inequality of 
\begin{equation}\label{EqInequality1}
\int d\varepsilon^\prime D(\varepsilon^\prime)
\frac{ \left[x+ S_1(\varepsilon,\varepsilon^\prime) \right]^2 }{
S_1^2(\varepsilon,\varepsilon^\prime) +S_2^2(\varepsilon)} >0,
\end{equation}
or 
$Y_0(\varepsilon) x^2 + 2 Y_1(\varepsilon) x + Y_2(\varepsilon) >0$
is satisfied for any real $x$, it follows that
\begin{equation}\label{EqAppendixYYY}
Y_1^2(\varepsilon)-Y_0(\varepsilon)Y_2(\varepsilon) < 0.
\end{equation}
The inequality (\ref{EqImp}), $\Delta(\varepsilon) \ge
\mbox{Re}\lambda^2 \Gamma(\varepsilon + i0)$, holds.

\section{Fermi-liquid theory for a spin liquid in the Heisenberg model}
\label{SecSpin}
\subsection{Localized spin in $s$-$d$ model}
\label{SecSpin1}
The FL theory for the $s$-$d$ model, which is a prototype of the FL theory for the Heisenberg model, is first studied by a different approach from Nozi\`{e}res'. \cite{nozieres} 
In the $s$-$d$ limit,
which is defined by $\tilde{U}/|V|^2\rightarrow+\infty$ with $2(\epsilon_d -\tilde{\mu})+ \tilde{U}=0$ and $J_{s\mbox{-}d} = -2|V|^2/\tilde{U}$ kept constant, the Anderson model (\ref{EqAnderson}), but with constant $V({\bf k})=V$ and 
\begin{equation}\label{EqConstDc}
\frac1{N_A} \sum_{\bf k} 
\delta[\varepsilon + \tilde{\mu} - E_c({\bf k})] = D_c(0),
\end{equation}
is reduced to the $s$-$d$ model:
\begin{equation}\label{EqSD}
{\cal H}_{s\mbox{-}d} =
- J_{s\mbox{-}d} \sum_{\tau\tau^\prime}
\bigl({\bf S}\cdot {\bm \sigma}^{\tau\tau^\prime}\bigr)
c_{0\tau}^\dag c_{0\tau^\prime} 
+ \sum_{{\bf k}\sigma} E_c({\bf k}) 
c_{{\bf k}\sigma}^\dag c_{{\bf k}\sigma},
\end{equation}
with ${\bf S}$ a localized spin with $S=1/2$ at the $0\hskip1pt$th site and
$c_{0\tau}=(1/\sqrt{N})\sum_{\bf k} c_{{\bf k}\tau}$.
A constant term of $\epsilon_a$ is ignored in Eq.~(\ref{EqSD}).

It is straightforward to extend the FL theory for the Anderson model \cite{yosida-yamada} to the $s$-$d$ limit.
The Green functions for $d$ electrons and conduction electrons in the Anderson model are given, respectively, by
\begin{equation}\label{EqG-SD}
\tilde{G}_\sigma(\varepsilon +i0)=
\frac1{\tilde{\phi}_\gamma}\left[\frac1{
\varepsilon + i \Delta^* } 
+ O(\varepsilon^2)\right],
\end{equation}
and
\begin{eqnarray}\label{EqScatt1}
G_{{\bf k}{\bf k}^\prime\sigma}(\varepsilon \!+\! i0) &=&
g_{{\bf k}\sigma}(\varepsilon \!+\! i0)
+ g_{{\bf k}\sigma}(\varepsilon \!+\! i0)
g_{{\bf k}^\prime\sigma}(\varepsilon \!+\! i0)
\nonumber \\ && \times
\frac{\Delta^*}{\pi D_c(0)}\left[\frac{1}{
\varepsilon + i \Delta^* } +
O(\varepsilon^2) \right],
\end{eqnarray}
for $|\varepsilon|\ll k_{\rm B}T_{\rm K}$, with
\begin{equation}\label{EqDelta*TK}
\Delta^* = \frac{1}{\tilde{\phi}_\gamma}\pi |V|^2 D_c(0)
= \frac{4}{\pi} k_{\rm B} T_{\rm K},
\end{equation}
and
$g_{{\bf k}\sigma}(\varepsilon +i0)=
1/[\varepsilon + \tilde{\mu} -E_c({\bf k}) + i0]$.
%
In Eq.~(\ref{EqDelta*TK}), Eq.~(\ref{EqDefTK}) is used.
Since $T_{\rm K}$ is nonzero and finite in the $s$-$d$ limit, 
\begin{equation}\label{EqPhiG-U}
\tilde{\phi}_\gamma \propto |V|^2 \propto \tilde{U},
\end{equation}
is satisfied there. 
In the $s$-$d$ limit or in the limit of $\tilde{\phi}_\gamma\rightarrow +\infty$, the Green function of $d$ electrons is vanishing, as is shown in Eq.~(\ref{EqG-SD}), but the fermionic spectrum of single-particle excitations for $d$ electrons is still well defined, as is shown in Eq.~(\ref{EqScatt1}).

In the $s$-$d$ model, $d$ electrons are exactly localized and they carry a localized spin. It is trivial that
the single-particle or fermionic Green function of the localized spin can never be defined. However, the fermionic spectrum, which describes scatterings of conduction electrons by the localized spin, is defined by Eq.~(\ref{EqScatt1}).
Physical properties of the localized spin in the $s$-$d$ model can be described by the fermionic spectrum according to the FL relation. For example, the specific heat is given by $C= \gamma T +\cdots$ at $T \ll k_{\rm B}T_{\rm K}$, with $\gamma$ given by Eq.~(\ref{EqFLR-G}). This is simply the FL theory for the $s$-$d$ model by Nozi\`{e}res.\cite{nozieres}

The number of $d$ electrons that carry a localized spin is a conserved quantity in the $s$-$d$ model but the number of $d$ electrons is not in the Anderson model, i.e., local gauge symmetry exists in the $s$-$d$ model but it does not in the Anderson model.
It should be noted that, however, the adiabatic continuity \cite{AndersonText} holds between the FL in the $s$-$d$ model and the FL in the Anderson model. \cite{comI-L}
It should also be noted that the Kondo peak, to which the Gutzwiller band in the Hubbard model under S\hskip2pt$^3$A or DMFT corresponds, appears in the fermionic spectrum even if no $d$ electron can be added or removed in the $s$-$d$ model but when an electron can be added or removed in the conduction band, as is shown by Eq.~(\ref{EqScatt1}).

\subsection{Spin liquid in the Heisenberg model}
\label{SecSpin2}
In the Heisenberg limit, which is defined by $U/|t|\rightarrow+\infty$ with $2(\epsilon_a-\mu) + U=0$ and $J=-4t^2/U$ kept constant, the Hubbard model with the electron reservoir, which is defined by Eq.~(\ref{EqGrandH}), is reduced to the Heisenberg model with a thermal reservoir:
\begin{eqnarray}\label{EqHeisenbergH}
\bar{\cal H}_{\rm H} \hskip-2pt &=& \hskip-2pt
-\frac1{2}J \sum_{\left<ij\right>} 
\left({\bf S}_i \cdot {\bf S}_j \right)
+ \sum_{{\bf k}\sigma} (E_b({\bf k}) -\mu)
b_{{\bf k}\sigma}^\dag b_{{\bf k}\sigma} ,
\nonumber \\ && 
- \lambda^2 J_r 
\!\sum_{i\in {\cal R}} \sum_{\tau\tau^\prime}
\bigl({\bf S}_i \cdot {\bm \sigma}^{\tau\tau^\prime}\bigr)
b_{i\tau}^\dag b_{i\tau^\prime} 
\end{eqnarray}
with ${\bf S}_i$ a localized spin with $S=1/2$ at the $i\hskip1pt$th site and $J_r = -2|v|^2/U$. A constant term of $ N_{\rm c} (\epsilon_a-\mu)$ is ignored in Eq.~(\ref{EqHeisenbergH}). 
First, two dimensions and higher are assumed.
When the RVB mechanism is only considered beyond S$^3$A or DMFT, the ground state is the FL even in the Heisenberg limit provided that the perturbation $\lambda^2J_r$ from the thermal reservoir is weak enough. 
Then, it is assumed that
\begin{subequations}\label{EqV2-0}
\begin{equation}\label{EqV2-1}
\lim_{U/|t|\rightarrow+\infty} |v|^2/U < +\infty .
\end{equation}
and
\begin{equation}\label{EqV2-2}
\lim_{U/|t|\rightarrow+\infty}|v|^2/\tilde{\phi}_\gamma < +\infty ,
\end{equation}
\end{subequations}
with $\tilde{\phi}_\gamma$ one of the expansion coefficients for the self-energy renormalized by the Fock-type term; the model (\ref{EqHeisenbergH}) and quasi-particles are well defined when Eq.~(\ref{EqV2-0}) is satisfied. 
It is also assumed that the FL is totally stabilized by the RVB mechanism, i.e.,
it is assumed according to Eq.~(\ref{EqW*}) and $W \propto |t| \propto \sqrt{U}$ that $\tilde{\phi}_\gamma$ satisfies
\begin{equation}\label{EqV2-3}
\lim_{U/|t|\rightarrow+\infty}\sqrt{U}/\tilde{\phi}_\gamma =0 .
\end{equation}

The Green functions for electrons in the Hubbard model and the reservoir averaged over the ensemble are given, respectively, by
\begin{equation}\label{EqGrennVanish}
G_\sigma(\varepsilon \!+\! i0, {\bf k})=
\frac1{\tilde{\phi}_\gamma}\left[\frac1{
\varepsilon-\xi({\bf k})
+i \lambda^2 \Gamma^* (\varepsilon\!+\!i0)} + O(\varepsilon^2)\right],
\end{equation}
and
\begin{eqnarray}\label{EqFL-Heisenberg}
G_{b\sigma}(\varepsilon \!+\! i0, {\bf k}) \hskip-3pt &=& \hskip-3pt
g_{b\sigma}(\varepsilon \!+\! i0, {\bf k})
+ n_{\rm h} \lambda^2 \frac{|v|^2}{\tilde{\phi}_\gamma}
 g_{b\sigma}^2(\varepsilon \!+\! i0, {\bf k})
\nonumber \\ && \hskip-20pt \times 
\left[\frac{1}{\varepsilon \!-\! \xi({\bf k}) \!+\! i \lambda^2 \Gamma^* (\varepsilon \!+\! i0) } \!+\! O(\varepsilon^2)\right] , 
\end{eqnarray}
for $|\varepsilon|\ll |c_J J|$, with 
$g_{b\sigma}(i\varepsilon_n, {\bf k})=
1/[i\varepsilon_n +\mu -E_b({\bf k})]$,
%
$\xi({\bf k})$ given by Eq.~(\ref{EqDisperBeyond}), and 
\begin{equation}\label{EqGamStar}
\Gamma^* (\varepsilon + i0) = 
i n_{\rm h} \frac{|v|^2}{\tilde{\phi}_\gamma} 
\frac1{N_{\rm c}}\sum_{\bf k}
g_{b\sigma}(\varepsilon +i0, {\bf k}).
\end{equation}
In the Heisenberg limit,
the Green function for electrons is vanishing.
Provided that Eq.~(\ref{EqV2-2}) is satisfied, however, quasi-particle excitations are well defined.
 
The conductivity in the Heisenberg limit is given by
\begin{eqnarray}\label{EqConductivity}
\sigma_{xx}(0) &\propto& \frac{e^2}{N_{\rm c}}
\sum_{\bf k}
\frac{\partial\phantom{k_x}}{\partial k_x} E({\bf k}) \frac{\partial\phantom{k_x}}{\partial k_x} \left[E({\bf k}) \!+\! \Delta\Sigma_\sigma({\bf k})\right]
\nonumber \\ && \qquad \times
\left[\mbox{Im}G_\sigma(+i0, {\bf k}) \right]^2
\nonumber \\ &\propto&
\frac{|t|}{|v|^2\lambda^2}ne^2|J|,
\end{eqnarray}
where the vertex correction consistent with the Fock-type self-energy is included. 
The asymptotic behavior of $J_r$ or $|v|^2$ should be properly assumed to satisfy Eq.~(\ref{EqV2-0}).
Assume that
$\tilde{\phi}_\gamma \propto U^\kappa $,
%
where $\kappa>1/2$ is required by Eq.~(\ref{EqV2-3}).
If $1/2<\kappa<1$, $|v|^2 \propto U^\kappa$ is assumed; $\Gamma^*(\varepsilon+i0)$ is nonzero and finite and $J_r$ is vanishing. 
If $\kappa=1$, $|v|^2 \propto U$ is assumed; both of $\Gamma^*(\varepsilon+i0)$ and $J_r$ are nonzero and finite.
If $\kappa>1$, $|v|^2 \propto U$ is assumed; $\Gamma^*(\varepsilon+i0)$ is vanishing and $J_r$ is nonzero and finite.
In either case, the conductivity is vanishing because $|t|/(|v|^2\lambda^2)\rightarrow 0$ in the Heisenberg limit followed by the limit of $\lambda\rightarrow 0$.
 
Assume that $\lambda^2$ is small but nonzero.
When $1/2< \kappa< 1$, $\lambda^2 \Gamma^*(\varepsilon+i0)$ is nonzero for vanishing $\lambda^2 J_r$, which means that there is a singularity at $\lambda^2 J_r=0$. When $|v|^2 \propto U$ is assumed, for example, $\lambda^2 \Gamma^* (\varepsilon + i0)$ is diverging for nonzero and finite $\lambda^2 J_r$, even if $\lambda^2 J_r$ is infinitesimally small. The ground state is completely disordered due to an infinitesimally small $\lambda^2J_r$. This is unreasonable.
When $\kappa>1$, $\lambda^2 \Gamma^* (\varepsilon + i0)$ is vanishing for nonzero and finite $\lambda^2 J_r$, even if $\lambda^2 J_r$ is large. A finitely large perturbation has no effect on the ground state. This is also unreasonable. 
We speculate that the truth is $\kappa=1$ or 
$\tilde{\phi}_\gamma \propto U$, as it is in the $s$-$d$ limit for the Anderson model.

When the RVB mechanism is only considered beyond S$^3$A or DMFT 
within the restricted Hilbert subspace, the ground state in the Heisenberg limit is a spin liquid, where
the Green function of electrons and
the conductivity are vanishing but the fermionic spectrum $\xi({\bf k})$ of almost localized electrons is defined. 
It is probable that, under the corresponding approximation within the restricted Hilbert subspace, the spin liquid is also the ground state of the Heisenberg model, where the single-particle Green function of localized spins can never be defined and the conductivity exactly vanishes. Although $\xi({\bf k})$ related to localized spins is also defined, it does not exist in the Heisenberg model.
Physical properties of the spin liquid can be described by $\xi({\bf k})$ according to the FL relation. \cite{Luttinger1,Luttinger2} For example, the specific heat per unit cell is given by $C= \gamma T +\cdots$ at $T \ll |c_JJ|$, with $\gamma$ given by Eq.~(\ref{EqFLR-G}). This is simply the FL theory for the Heisenberg model.

The FS exists in the Hubbard model but it does not in the Heisenberg model. 
In general, no change of symmetry occurs between a metal, in which the FS exists, and an insulator, in which no FS exists, if no order parameter is involved; the breaking of local gauge symmetry may occur. \cite{comI-L}
The analysis in this Appendix implies that the adiabatic continuity \cite{AndersonText} holds between the FL in the Hubbard model and the spin liquid in the Heisenberg model, i.e., the spin liquid is simply the FL. 
We propose that the insulating RVB state \cite{fazekas} is the spin liquid and, therefore, the metallic RVB state \cite{RVB} is simply the FL.

In one dimension, terms proportional to $\varepsilon \ln |\varepsilon|$ appear in the multi-site self-energy at $T=0$~K. Even in this case, it is possible to describe the Green functions by Eqs.~(\ref{EqGrennVanish}) and (\ref{EqFL-Heisenberg}), where each of $1/[\varepsilon -\xi({\bf k})+i \lambda^2 \Gamma^* (\varepsilon\!+\!i0)]$ is replaced by
$1/[\varepsilon + c J \varepsilon \ln |\varepsilon| -\xi({\bf k})+i \lambda^2 \Gamma^* (\varepsilon\!+\!i0)]$, with $c$ a numerical constant. It is straightforward to extend the above analysis to one dimension. The extended analysis implies that the adiabatic continuity \cite{AndersonText} holds between the Tomonaga-Luttinger liquid in the Hubbard model and Bonner and Fishers's spin liquid \cite{bonner} in the Heisenberg model. 

The Tomonaga-Luttinger liquid and Lieb and Wu's spin liquid seem to be of the same symmetry as each other.
It is interesting to examine whether the adiabatic continuity \cite{AndersonText} holds between them.
If the adiabatic continuity holds between the Tomonaga-Luttinger liquid and Bonner and Fishers's spin liquid, as is discussed above, it is probable that the adiabatic continuity also holds between the Tomonaga-Luttinger liquid and Lieb and Wu's spin liquid.
When $\lambda=\pm 0$, $N$ is a non-conserved quantity. When $U/|t|$ is finite and the chemical potential is continuos at $n=1$ as a function of $n$, a state for $n=1$ is, in a certain sense, an average over states with $N =N_{\rm c} +\Delta N$, with $\Delta N = 0$, $\pm 1$, $ \pm 2$, $\cdots$, in the thermodynamic limit.
One may argue that if the ground states for $n\rightarrow 1\pm0$ are the same as each other the ground state for $n=1$ must be the same as those for $n\rightarrow 1\pm0$. If this argument is relevant and the adiabatic continuity holds between the Tomonaga-Luttinger liquid and Lieb and Wu's spin liquid, it is possible that the ground state for $\lambda=\pm 0$, $n=1$, and finite $U/|t|$ in one dimension is simply the Tomonaga-Luttinger liquid.

\section{Single-particle or fermionic excitations}
\label{SecAppendixGutzwiller}
\subsection{Grand canonical ensemble}
\label{SecAppendixGrand}
One of the purposes of this Appendix is to study a general feature of single-particle or fermionic excitations in the grand canonical and canonical ensembles under an assumption that $d\ge 1$, $n\simeq 1$ or $N\simeq N_{\rm c}$, and $1\ll U/|t|< +\infty$. It is shown that fermionic excitations, which themselves are never {\it observables}, are different between the two ensembles. Then, it is shown that the well-known physical picture for the Mott transition, which is one in the canonical ensemble, 
is never relevant to explain the Mott transition.

First, consider the Hubbard model in the grand canonical ensemble or $\bar{\cal H}$ defined by Eq.~(\ref{EqGrandH}). 
The eigen-equation is given by 
\begin{equation}
\bar{\cal H} \big|\bar{N};\alpha \bigr> = 
\bar{E}_{\bar{N};\alpha} \big|\bar{N};\alpha \bigr> ,
\end{equation}
with $\bar{N}$ the number of total electrons and $\alpha$ a quantum number for an eigen-state of $\bar{\cal H}$.
The ground state for $\mu$ that corresponds to $n=1$ is denoted by $\big|\bar{N}_g;g\bigr>$.
The thermodynamic limit of $\bar{N}_g\rightarrow +\infty$ is assumed; $\mu$'s are the same as each other among $\bar{N}=\bar{N}_g$, $\bar{N}_g\pm1$, $\cdots$.

When $|t|/U\rightarrow +0$, it follows that $(\mu - \epsilon_a)/|t| \rightarrow +\infty$ and $(\epsilon_a + U -\mu)/|t|\rightarrow +\infty$.
All the unit cells in the Hubbard model are singly occupied
in either of the ground states for $\bar{N}=\bar{N}_g$, $\bar{N}_g\pm 1$, $\cdots$;
{\it extra electrons go to the reservoir or deficient ones come
from it}. They are degenerate with each other such that
$\bar{E}_{\bar{N}_g;g}$ $=$ $\bar{E}_{\bar{N}_g\pm 1;g}$ $=\cdots$,
which is simply denoted by $\bar{E}_g$.
Eigen-states are classified according to the numbers 
of doubly occupied and empty unit cells in the
Hubbard model, which are denoted by 
$d_{\bar{N};\alpha}$ and $e_{\bar{N};\alpha}$, respectively:
\begin{equation}\label{EqDegeneracy}
\bar{E}_{\bar{N};\alpha} = 
\bar{E}_g+ d_{\bar{N};\alpha}\left(\epsilon_a + U - \mu\right)
+e_{\bar{N};\alpha}\left(\mu - \epsilon_a\right) ,
\end{equation}
for ${\bar N}=N_g$, $N_g \pm 1$, $\cdots$. 
They are degenerate with each other when
their $d_{\bar{N};\alpha}$ and $e_{\bar{N};\alpha}$ are the same 
as each other, even if their $\bar{N}$'s are different from each other.

When $0<|t|/U\ll 1$, it follows that $\mu - \epsilon_a \simeq U/2 \gg |t|$ and $\epsilon_a + U - \mu \simeq U/2 \gg |t|$. Eigenstate can be still classified according to 
$d_{\bar{N};\alpha}$ and $e_{\bar{N};\alpha}$.
The degeneracy (\ref{EqDegeneracy}) in 
the ground and low-lying states with $d_{\bar{N};\alpha}=0$ and $e_{\bar{N};\alpha}=0$ is lifted by second-order perturbation in $t$: 
\begin{equation}\label{EqAppGutWidth}
\bar{E}_{\bar{N};\alpha} = \bar{E}_g \pm O(|J|),
\end{equation}
for ${\bar N}=N_g$, $N_g \pm 1$, $\cdots$, with $J$ the superexchange interaction.
It should be noted that the ground states for ${\bar N}=N_g$, $N_g \pm 1$, $\cdots$ are still degenerate with each other.
The $O(|J|)$ term in Eq.~(\ref{EqAppGutWidth}) implies that each of the ground states is a singlet stabilized by the RVB mechanism.
Since a doubly-occupied or empty site is {\it itinerant}, the degeneracy (\ref{EqDegeneracy}) in excited states with $d_{\bar{N};\alpha}\ge 1$ or $e_{\bar{N};\alpha}\ge 1$ is lifted by first-order perturbation in $t$:
\begin{equation}\label{EqAppUHB-LHB}
\bar{E}_{\bar{N};\alpha} = \bar{E}_g + d_{\bar{N};\alpha} 
\left(\epsilon_a \!+\! U \!-\! \mu\right)
+ e_{\bar{N};\alpha} 
\left(\mu \!-\! \epsilon_a\right) \pm O(W) ,
\end{equation}
for ${\bar N}=N_g$, $N_g \pm 1$, $\cdots$.

When $T=0$~K, DOS defined by Eq.~(\ref{EqRhoSSA}) is given by
\begin{eqnarray}\label{EqAppRhoGrand}
\rho(\varepsilon) &=& 
\frac1{N_{\rm c}}\sum_{{\bf k}\alpha}\Bigl[ A_{{\bf k}\alpha}^{+}
\delta \bigl(\varepsilon - \bar{E}_{\bar{N}_g+1;\alpha} 
+ \bar{E}_{\bar{N}_g;g}\bigr)
\nonumber \\ &&
+ A_{{\bf k}\alpha}^{-}
\delta\bigl(\varepsilon + \bar{E}_{\bar{N}_g-1;\alpha} 
- \bar{E}_{\bar{N}_g;g}\bigr) 
\Bigr], 
\end{eqnarray}
with $N_{\rm c}$ the number of unit cells of the Hubbard model,
\begin{subequations}\label{EqAppApm}
\begin{equation}
A_{{\bf k}\alpha}^{+} =
\bigl| \bigl< \bar{N}_g+ 1;\alpha\bigl| a_{{\bf k}\sigma}^\dag 
\bigr| \bar{N}_g;g \bigr>\bigr|^2 ,
\end{equation}
and
\begin{equation}
A_{{\bf k}\alpha}^{-} =
\left| \left< \bar{N}_g-1;\alpha\big| a_{{\bf k}\sigma}
\big| \bar{N}_g;g \right>\right|^2.
\end{equation}
\end{subequations}
In the grand canonical ensemble, the fermionic spectrum $\rho(\varepsilon)$ is related to response functions or {\it observables}, which are bosonic, according to the FL relation.\cite{Luttinger1,Luttinger2} 

Low-energy excitations are possible from the ground or initial state $\left|\bar{N}_g;g\right>$ to low-lying {\it final} states $\left|\bar{N}_g \pm 1;\alpha\right>$ with $d_{\bar{N}_g\pm 1;\alpha}= 0$ and $e_{\bar{N}_g\pm 1;\alpha}= 0$. 
A narrow band, whose width is $O(|J|)$ according to Eq.~(\ref{EqAppGutWidth}), appears around $\varepsilon \simeq 0$ or around the chemical potential.
The nature of the band depends on that of the ground state.
When the ground state is the FL, the band is simply the Gutzwiller band renormalized by the RVB mechanism.
A band or structure also appears, or at least $|t|\rho(0)>0$, in the Tomonaga-Luttinger liquid. If the ground state is a spin liquid, $|t|\rho(0)=0$ but a structure must appear in $\rho(\varepsilon)$ at $|\varepsilon| \alt O(|J|)$.
It is also possible that a complete gap opens if an order parameter appears.

Even if the ground state is a spin liquid, the charge-spin separation occurs, or an order parameter appears, excitations as large as $U$ are possible from the ground state to final states with $d_{\bar{N}_g+1;\alpha}= 1$ and $e_{\bar{N}_g+1;\alpha}= 0$ and those with $d_{\bar{N}_g-1;\alpha}= 0$ and $e_{\bar{N}_g-1;\alpha}= 1$.
Two broad bands, whose widths are $O(W)$ according to Eqs.~(\ref{EqAppGutWidth}) and (\ref{EqAppUHB-LHB}), appear around $\varepsilon \simeq \pm U/2$ below and above the chemical potential. They are simply UHB and LHB. When $U\agt W$, the Hubbard gap or pseudo-gap opens for any $d\ge 1$ between UHB and LHB.

\subsection{Canonical ensemble}
Next, consider a Hubbard-like model in the canonical ensemble: 
\begin{equation}
{\cal H}_{U^\prime}={\cal H} + U^\prime \left({\cal N}-N_{\rm c}\right)^2, 
\end{equation}
with ${\cal H}$ the Hubbard model (\ref{EqHubbard}) and ${\cal N}$ the number operator (\ref{EqNumber}).
Since ${\cal N}$ is a conserved quantity, none of observables depend on $U^\prime$ since $U^\prime \left({\cal N}-N_{\rm c}\right)^2$ is simply a constant of $U^\prime \left(N-N_{\rm c}\right)^2$.
The eigen-equation is given by 
\begin{equation}
{\cal H}_{U^\prime} \big| N ;\alpha\bigr> = E_{N;\alpha}\big| N;\alpha\bigr> .
\end{equation}
Since almost all the unit cells are singly occupied in the ground and low-lying states for $N=N_{\rm c}$, it follows that
\begin{equation}\label{EqCanonical1}
E_{N_{\rm c};\alpha} =  N_{\rm c} \epsilon_a \pm O(|J|) ,
\end{equation}
by second-order perturbation in $t$. When a pair of empty and doubly occupied site appear in an excited state, 
%
\begin{equation}\label{EqCanoExcite}
E_{N_{\rm c};\alpha} =  N_{\rm c} \epsilon_a + U \pm O(W) ,
\end{equation}
by first-order perturbation in $t$.
{\it When an electron or a hole is added, it remains within the Hubbard model, which is sharp contrast with that it can escape from the Hubbard model in the grand canonical ensemble}. By first-order perturbation in $t$, 
\begin{subequations}\label{EqCanonical}
\begin{equation}\label{EqCanonical2}
E_{N_{\rm c}+1:\alpha} = U^\prime + (N_{\rm c}+1)\epsilon_a +U \pm O(W),
\end{equation}
and
\begin{equation}\label{EqCanonical3}
E_{N_{\rm c}-1:\alpha} = U^\prime +(N_{\rm c}-1)\epsilon_a \pm O(W),
\end{equation}
\end{subequations}
for the ground and low-lying states. 

The $O(|J|)$ term in Eq.~(\ref{EqCanonical1}) implies that the ground state for $N=N_{\rm c}$ is a singlet for even $N_{\rm c}$ or a doublet for odd $N_{\rm c}$ stabilized by the RVB mechanism.
It also implies that spin fluctuations with energy scale $O(|J|)$ are developed.
When no charge-spin separation occurs, spin fluctuations inevitably couple with charge fluctuations. The existence of coupled charge-spin fluctuations implies that the ground state is a metal.
When the charge-spin separation occurs, the ground state may be a spin liquid or a metal.
The $\pm O(W)$ term in Eq.~(\ref{EqCanonical}), which is due to the itineracy of a doubly-occupied or empty site, implies that either of the ground states for $N=N_{\rm c} \pm 1$ is a metal.
A phase diagram speculated from this argument is consistent with that for one dimension according to the Bethe-ansatz solution, where the ground state for finite $U$ is a spin liquid for $N=N_{\rm c}$ but is a metal for $N\ne N_{\rm c}$.
The argument implies that the ground state for finite $U$ is a metal for any $N$ in two dimensions and higher, if no charge-spin separation occurs or no order parameter appears.

Consider the conductivity $\sigma_{xx}(\omega)$ for $N=N_{\rm c}$. It is mainly determined by two-particle excitations. When a pair of single-particle excitations, a particle and a hole, are bound, i.e., almost all unit cells are singly occupied in a pair-excited state, it follows that
\begin{equation}\label{EqPair1}
E_{N_{\rm c};\alpha}-E_{N_{\rm c};g} = O(|J|),
\end{equation}
with the ground state denoted by $\alpha=g$.
When a particle and a hole are not bound in an excited state, 
\begin{equation}\label{EqPair2}
E_{N_{\rm c};\alpha}-E_{N_{\rm c};g} = U \pm O(W).
\end{equation}
None of pair excitations depend on $U^\prime$.
Two structures can appear in $\mbox{Re}\hskip2pt \sigma_{xx}(\omega)$: a low-energy one around $\omega=O(|J|)$
and a high-energy one around $\omega=U$.
If the ground state is a spin liquid, a small gap opens within the low-energy structure. The small gap is $O(|J|)$ or smaller than $O(|J|)$. 
If the ground state is a metal, no gap opens.
If the ground state is the FL, in particular, the low-energy structure is simply the Drude term due to quasi-particles, whose band-width is $O(|J|)$.
The high-energy structure corresponds to a pair excitation between the Hubbard gap.
The analysis on the conductivity, which is an observable, in the canonical ensemble is consistent with that on single-particle excitations in the grand canonical ensemble, which are related to observables according to the FL relation.\cite{Luttinger1,Luttinger2} 

A fermionic spectrum is defined by
\begin{eqnarray}\label{EqAppRhoCanonical}
X_{\rho}(\varepsilon) &=& 
\frac1{N_{\rm c}}\sum_{{\bf k}\alpha}\Bigl[ B_{{\bf k}\alpha}^{+}
\delta \bigl(\varepsilon - E_{N_{\rm c}+1;\alpha} 
+ E_{N_{\rm c};g}\bigr)
\nonumber \\ &&
+ B_{{\bf k}\alpha}^{-}
\delta\bigl(\varepsilon + E_{N_{\rm c}-1;\alpha} 
- E_{N_{\rm c};g}\bigr) 
\Bigr], 
\end{eqnarray}
with 
\begin{subequations}
\begin{equation}
B_{{\bf k}\alpha}^{+} =
\bigl| \bigl< N_{\rm c}+ 1;\alpha\bigl| a_{{\bf k}\sigma}^\dag 
\bigr| N_{\rm c};g \bigr>\bigr|^2 ,
\end{equation}
and
\begin{equation}
B_{{\bf k}\alpha}^{-} =
\left| \left< N_{\rm c}-1;\alpha\big| a_{{\bf k}\sigma}
\big| N_{\rm c};g \right>\right|^2.
\end{equation}
\end{subequations}
A gap as large as $U+2U^\prime- O(W)$ opens in $X_{\rho}(\varepsilon)$.
When $U^\prime=0$, ${\cal H}_{U^\prime}$ is simply the Hubbard model ${\cal H}$.
One may argued 
that the Hubbard gap as large as $U-O(W)$ or $U-W$ opens in $X_{\rho}(\varepsilon)$ so that the ground state for $N=N_{\rm c}$ is the Mott insulator for $U \agt W$. This is simply the well-known physical argument or picture for the Mott transition.
However, the picture is never relevant. First of all, $X_{\rho}(\varepsilon)$ depends on $U^\prime$, which means that $X_{\rho}(\varepsilon)$ is not related to observables or, at least, it is not directly related to observables. 
When the ground state is a spin liquid, a gap in $X_{\rho}(\varepsilon)$ is different from the small gap in the conductivity $\mbox{Re}\hskip2pt \sigma_{xx}(\omega)$.
When the ground state is a metal, $X_{\rho}(\varepsilon)$ cannot describe observables of the metal, whose energy scale is $O(|J|)$. 

All the analyses and arguments in this paper show or imply that the Mott insulator is impossible when $U/|t|$ is finite, $J$ is nonzero, or the RVB mechanism is effective. An exception is the well-known physical picture.
The FL theories for the $s$-$d$ model and the Heisenberg model in Appendix~\ref{SecSpin} imply that a relevant fermionic spectrum, which is related to observables, can be defined even in the canonical ensemble when a thermal reservoir is explicitly considered, i.e., when an electron is added or removed in the thermal reservoir even if no electron is added or removed in the Hubbard model.


\begin{thebibliography}{99}
\bibitem{Hubbard3}
The Hubbard approximation is also formulated as the coherent potential approximation (CPA) for a disordered system where electrons with up and down spins are randomly distributed. This CPA theory implicitly assumes that the entropy is $k_{\rm B}\ln 2$ per unit cell for $n=1$.
\bibitem{comAttractive}
Relevant leading-order or $O(1)$ effects in $1/d$ are different when $U$ is repulsive or attractive. When $U$ is repulsive, they are single-site spin fluctuations and magnetic order or SDW.
When $U$ is attractive, they are single-site charge fluctuations, CDW, isotropic $s$-wave or BCS superconducting single-site fluctuations, and BCS superconductivity.
The single-site fluctuations are rigorously considered under S$^3$A, DMFT, or DCPA. However, ordered states of SDW, CDW, and BCS superconductivity and corresponding susceptibilities can only be treated beyond it.
\bibitem{comHubbard}
The Mott insulator in the Hubbard approximation is also characterized by the self-energy of electrons having a pole on the real axis, so that the Hubbard gap is a complete gap between UHB and LHB.
Even if a self-energy function that has a pole on the real axis is assumed in Eq.~(\ref{EqMappingCondition2}), Eq.~(\ref{EqImp}) or the FS condition (\ref{EqFS}) is satisfied. Therefore, such a singular self-energy function can never be a self-consistent solution under S$^3$A, DMFT, or DCPA.
\bibitem{comMappingBeyondSSA} 
The mapping to the Anderson model is parallel with that in Sec.~\ref{SecMapping} when DOS defined by $E({\bf k})+\Delta \Sigma_\sigma({\bf k})$ is used instead of $D(\varepsilon)$ defined by $E({\bf k})$. Since the FS condition (\ref{EqFS}) is satisfied, the renormalized expansion coefficients by the Fock-type term are also real and finite.
\bibitem{Takahashi2}
One may argue according to Takahashi's analysis \cite{Takahashi1} that the entropy of Lieb and Wu's insulator is nonzero, even if $N$ is even. However, if there is no other solution besides the Bethe-ansatz solution, the ground state for finite $U/|t|$ is a singlet for even $N$ or a doublet for odd $N$.
\bibitem{comSeparation}
When excitation energies or $k_{\rm B}T$ are higher than $O(U)$, the charge-spin separation never occurs even in one dimension. For example, it never occurs in pair excitations between UHB and LHB. This fact implies that, when $U/|t|$ is finite, the separation is rigorous only in the limit of vanishing excitation energies and $T\rightarrow 0$~K.
\bibitem{comI-L} 
In general, no change of symmetry occurs, except local gauge symmetry, and the adiabatic continuity holds between itinerant and localized states, if no order parameter is involved.
In the Tomonaga-Luttinger liquid,
no symmetry change occurs between itinerant spins for finite $U/|t|$ and localized spins for $U/|t|=+\infty$.
Rigorously speaking, no mobility edge exists in the Anderson localization, but a sharp crossover occurs between itinerant and localized states.
\end{thebibliography}
\end{document}